\documentclass[11pt,letter]{article}
\usepackage{amsmath,amsthm,latexsym,amssymb,amsfonts,epsfig}

\makeatletter
\@addtoreset{equation}{section}
\makeatother

\newtheorem{Theorem}{Theorem}[section]

\def\be{\begin{equation}}
\def\ee{\end{equation}}
\def\ba{\begin{eqnarray}}
\def\ea{\end{eqnarray}}

\newcommand\nn{\nonumber}
\newcommand\q{\quad}
\def\a{{\cal A}}

\def\Nl{{\mathchoice
{\setbox0=\hbox{$\displaystyle\rm N$}\hbox{\hbox to0pt
{\kern0.4\wd0\vrule height0.9\ht0\hss}\box0}}
{\setbox0=\hbox{$\textstyle\rm N$}\hbox{\hbox to0pt
{\kern0.4\wd0\vrule height0.9\ht0\hss}\box0}}
{\setbox0=\hbox{$\scriptstyle\rm N$}\hbox{\hbox to0pt
{\kern0.4\wd0\vrule height0.9\ht0\hss}\box0}}
{\setbox0=\hbox{$\scriptscriptstyle\rm N$}\hbox{\hbox to0pt
{\kern0.4\wd0\vrule height0.9\ht0\hss}\box0}}}}
\def\Zl{{\mathchoice
{\setbox0=\hbox{$\displaystyle\rm Z$}\hbox{\hbox to0pt
{\kern0.4\wd0\vrule height0.9\ht0\hss}\box0}}
{\setbox0=\hbox{$\textstyle\rm Z$}\hbox{\hbox to0pt
{\kern0.4\wd0\vrule height0.9\ht0\hss}\box0}}
{\setbox0=\hbox{$\scriptstyle\rm Z$}\hbox{\hbox to0pt
{\kern0.4\wd0\vrule height0.9\ht0\hss}\box0}}
{\setbox0=\hbox{$\scriptscriptstyle\rm Z$}\hbox{\hbox to0pt
{\kern0.4\wd0\vrule height0.9\ht0\hss}\box0}}}}
\def\Ql{{\mathchoice
{\setbox0=\hbox{$\displaystyle\rm Q$}\hbox{\hbox to0pt
{\kern0.4\wd0\vrule height0.9\ht0\hss}\box0}}
{\setbox0=\hbox{$\textstyle\rm Q$}\hbox{\hbox to0pt
{\kern0.4\wd0\vrule height0.9\ht0\hss}\box0}}
{\setbox0=\hbox{$\scriptstyle\rm Q$}\hbox{\hbox to0pt
{\kern0.4\wd0\vrule height0.9\ht0\hss}\box0}}
{\setbox0=\hbox{$\scriptscriptstyle\rm Q$}\hbox{\hbox to0pt
{\kern0.4\wd0\vrule height0.9\ht0\hss}\box0}}}}
\def\Rl{{\mathchoice
{\setbox0=\hbox{$\displaystyle\rm R$}\hbox{\hbox to0pt
{\kern0.4\wd0\vrule height0.9\ht0\hss}\box0}}
{\setbox0=\hbox{$\textstyle\rm R$}\hbox{\hbox to0pt
{\kern0.4\wd0\vrule height0.9\ht0\hss}\box0}}
{\setbox0=\hbox{$\scriptstyle\rm R$}\hbox{\hbox to0pt
{\kern0.4\wd0\vrule height0.9\ht0\hss}\box0}}
{\setbox0=\hbox{$\scriptscriptstyle\rm R$}\hbox{\hbox to0pt
{\kern0.4\wd0\vrule height0.9\ht0\hss}\box0}}}}
\def\Cl{{\mathchoice
{\setbox0=\hbox{$\displaystyle\rm C$}\hbox{\hbox to0pt
{\kern0.4\wd0\vrule height0.9\ht0\hss}\box0}}
{\setbox0=\hbox{$\textstyle\rm C$}\hbox{\hbox to0pt
{\kern0.4\wd0\vrule height0.9\ht0\hss}\box0}}
{\setbox0=\hbox{$\scriptstyle\rm C$}\hbox{\hbox to0pt
{\kern0.4\wd0\vrule height0.9\ht0\hss}\box0}}
{\setbox0=\hbox{$\scriptscriptstyle\rm C$}\hbox{\hbox to0pt
{\kern0.4\wd0\vrule height0.9\ht0\hss}\box0}}}}
\def\Hl{{\mathchoice
{\setbox0=\hbox{$\displaystyle\rm H$}\hbox{\hbox to0pt
{\kern0.4\wd0\vrule height0.9\ht0\hss}\box0}}
{\setbox0=\hbox{$\textstyle\rm H$}\hbox{\hbox to0pt
{\kern0.4\wd0\vrule height0.9\ht0\hss}\box0}}
{\setbox0=\hbox{$\scriptstyle\rm H$}\hbox{\hbox to0pt
{\kern0.4\wd0\vrule height0.9\ht0\hss}\box0}}
{\setbox0=\hbox{$\scriptscriptstyle\rm H$}\hbox{\hbox to0pt
{\kern0.4\wd0\vrule height0.9\ht0\hss}\box0}}}}
\def\Ol{{\mathchoice
{\setbox0=\hbox{$\displaystyle\rm O$}\hbox{\hbox to0pt
{\kern0.4\wd0\vrule height0.9\ht0\hss}\box0}}
{\setbox0=\hbox{$\textstyle\rm O$}\hbox{\hbox to0pt
{\kern0.4\wd0\vrule height0.9\ht0\hss}\box0}}
{\setbox0=\hbox{$\scriptstyle\rm O$}\hbox{\hbox to0pt
{\kern0.4\wd0\vrule height0.9\ht0\hss}\box0}}
{\setbox0=\hbox{$\scriptscriptstyle\rm O$}\hbox{\hbox to0pt
{\kern0.4\wd0\vrule height0.9\ht0\hss}\box0}}}}
\newcommand{\ca}{\mathcal A}

\newcommand{\cc}{\mathcal C}
\newcommand{\cd}{\mathcal D}

\newcommand{\cg}{\mathcal G}

\newcommand{\ci}{\mathcal I}

\newcommand{\cm}{\mathcal M}
\newcommand{\cn}{\mathcal N}
\newcommand{\co}{\mathcal O}

\newcommand{\cs}{\mathcal S}
\newcommand{\ct}{\mathcal T}

  \newcommand{\Fc}{\mathfrak{C}}

\newcommand{\fg}{\mathfrak{g}}

  \newcommand{\Ft}{\mathfrak{T}}

\renewcommand{\a}{\alpha}
\renewcommand{\b}{\beta}
\newcommand{\eps}{\epsilon}
\newcommand{\veps}{\varepsilon}
\renewcommand{\t}{\tau}

\begin{document}

\title{Partial and Complete Observables for Hamiltonian Constrained Systems}
\author{B. Dittrich\thanks{dittrich@aei.mpg.de and bdittrich@perimeterinstitute.ca}, \\
 MPI f. Gravitationsphysik, Albert-Einstein-Institut, \\
 Am M\"uhlenberg 1, 14476 Golm near Potsdam, Germany \\
and \\
Perimeter Institute for Theoretical Physics \\
31 Caroline Street North, Waterloo, ON N2L 2Y5, Canada \\
}

\date{{\small Preprint AEI-2004-102}}

\maketitle

\begin{abstract}
We will pick up the concepts of partial and complete observables introduced by Rovelli in \cite{RovPartObs} in order to construct Dirac observables in gauge systems. We will generalize these ideas to an arbitrary number of gauge degrees of freedom. Different methods to calculate such Dirac observables are developed. For background independent field theories we will show that partial and complete observables can be related to Kucha\v{r}'s Bubble Time Formalism \cite{bubble}. Moreover one can define a non-trivial gauge action on the space of complete observables and also state the Poisson brackets of these functions.

Additionally we will investigate, whether it is possible to calculate Dirac observables starting with partially invariant partial observables, for instance functions, which are invariant under the spatial diffeomorphism group.
\end{abstract}

\section{Introduction}\label{intro}

This work is concerned with Hamiltonian systems with first class constraints, or in other words systems with gauge degrees of freedom. Examples for such systems are provided by background independent theories, such as general relativity. For such systems one has not only to solve the constraints but also to find the gauge independent degrees of freedom, that is Dirac observables. These Dirac observables are crucial for the canonical quantization of first class systems, since only Dirac operators can be promoted into operators on the physical Hilbert space. Moreover one can see quantization as a process to look for a representation of the observable algebra on a Hilbert space. Hence it is important to know the properties of the observable algebra.

One of the outstanding problems towards a quantum theory of gravity is, that there are no explicit Dirac observables known  (with the exception of the ten ADM charges for the asymptotic flat case and a Dirac observable which assumes only a few discrete values, see \cite{jacobson}). Furthermore the connection between (spacetime-) diffeomorphism invariant functions and Dirac observables is not very clear. This coheres with the problem to represent the diffeomorphism group on the canonical phase space, see \cite{isham2}. There are some attempts to bypass these problems by enlarging the phase space in order to obtain a representation of the diffeomorphism group and a theory which is explicitly covariant with respect to this group, see for instance \cite{savvidou,kucharstring}.

In this work we will adhere to the canonical phase space and pick up an idea of Rovelli in \cite{RovPartObs} for the construction of Dirac observables. For systems with one constraint the idea works in the following way: Assume that the system is totally constrained so that the constraint generates the time evolution (which is then considered as a gauge transformation). Use a phase space function $T$, which is not a Dirac observable, as a clock which ``measures'' the time flow, i.e. the gauge transformation. Consider another phase space function $f$ and calculate the value of $f$ ``at the time'' at which $T$ assumes the value $\tau$. Since the value of $f$ at a fixed time $\tau$ does not change with time, the result will be time independent, i.e. a Dirac observable. Moreover varying $\tau$ gives a one-parameter family of Dirac observables. Following Rovelli we will call $T$ and $f$ partial observables and the one-parameter family of Dirac observables complete observables. 

In section \ref{prelim} we summarize the aspects of first class constraint systems important for this work and explain our notations. 

Section \ref{oneconstr} reviews partial and complete observables for systems with one constraint and proves that complete observables are Dirac observables ( if certain assumptions are satisfied).

We will develop the concepts of complete observables for canonical systems with an arbitrary (finite) number of constraints in section \ref{manyconstr}. In section \ref{PDE} we write down a system of partial differential equations for complete observables and give a formal solution as a series in powers of the form $(\tau-T)$. Section \ref{abelian} shows, that a key point in the construction is to indroduce a set of new constraints, where the associated flows commute at least on the constraint hypersurface.

In section \ref{gauge} we show that the concepts of partial and complete observables can also be understood as the gauge invariant extensions of gauge restricted functions, see \cite{henneaux}. This viewpoint enables us to calulate the Poisson brackets between complete observables. Moreover we can define a non-trivial action of gauge transformations on the set of complete observables (acting on the parameters $\tau$).

Section \ref{fields} considers (canonical) field theories, having infinitly many constraints. It will turn out (section \ref{bubble}) that the concepts of complete and partial observables are related to Kuchar's Bubble Time formalism \cite{bubble}: To define a complete observable we will need infinitly many clocks which describe the embedding of the spatial hypersurface into the space-time manifold. A complete observable is then a phase space function evaluated on an embedding which is fixed by prescribing certain values for the infinitly many clock variables. Moreover it is possible to define an (non-trivial) action of gauge transformations on complete observables. 

Furthermore we will discuss whether it is possible to calculate complete observables starting from spatially diffeomorphism invariant partial observables. More generally we will consider in section \ref{partinv} partial observables, which are invariant under a subalgebra of the constraints. As will be shown, this reduces the number of equations to solve, considerably.

We will end with a discussion in section \ref{discuss}.

\section{Preliminaries and Notation}\label{prelim}

In all sections but section \ref{fields} we will consider finite dimensional phase spaces. Here a phase space is a $2p$-dimensional smooth manifold $\cm$ with a non-degenrate Poissonbracket $\{\cdot,\cdot\}$, such that there are canonical coordinates $(q_a,p_a,\,a=1,\ldots,p)$ with
\ba
\{q_a,p_b\}=\delta_{ab}  \q .
\ea
$\cc^\infty(\cm)$ is the space of smooth phase space functions and throughout this article we will work in the category of smooth function, also if this is not explicitely mentioned. (However this is not always the case in the examples.) We will denote phase space points by $x=(q_a,p_a)$.  

The subject of this article are systems with first class constraints or in other words gauge systems. In this section we will just mention the facts relevant for this article, a more detailed introduction can be found for example in \cite{henneaux}.

First class constrained systems are characterized by the fact that the admissible initial data are restricted to a submanifold of the phase space $\cm$ which is called the constraint hypersurface $\cc$. This hypersurface is described by the vanishing of $n<p$ constraints $C_j,\,j=1,\ldots,n$, which are functions on $\cm$. 

We will always assume that the constraints are algebraically independent, i.e. that the maximal submanifold on which they vanish simultanously is $(2p-n)$-dimensional, where $2p$ is the dimension of the phase space $\cm$.  

The first class property means that
\be \label{1pre1}
\{C_i,C_j\}=f_{ijk} \, C_k 
\ee
where the $f_{ijk}$ are called structure constants if they are independent from the canonical variables and otherwise structure functions. For such first class systems the canonical analysis results in a Hamiltonian $H$ which is only uniquely defined on the constraint hypersurface $\cc$:
\be \label{1pre2}
H=h+\sum_{j=1}^n \lambda_j C_j
\ee
where $\lambda_j$ are arbitrary (smooth) phase space functions. Here $h$ is a function which either does not identically vanish on $\cc$ or is identical to zero. In the latter case the Hamiltonian is a sum of constraints and the system is therefore called totally constrained.

Because of the arbitrariness of the functions $\lambda_j$ in the Hamiltonian (\ref{1pre2}) the dynamics of the system cannot be unique. To account for this non-uniqueness one postulates
that different phase space points $x_1,x_2$ describe the same physical state if they are connected by a gauge transformation.

Here a gauge transformation is a transformation which is generated by the constraints $C_j$. To explain this in more detail we will introduce the notion of a phase space flow $\a_C^t$ generated by a smooth phase space function $C$. Firstly we note that to every phase space function $C$ one can associate a so called Hamiltonian vector field $\chi_C$ defined by the condition
\ba
\{C,f\}=\chi_C(f) 
\ea
which has to be satisfied for arbitrary smooth function $f$. Here $=\chi_C(f)$ denotes the action of the vector field $\chi_C$ on the function $f$. Then we can define the flow $\a^t_C(x)$ of a phase space point $x$ by demanding that for the tangent of the curve $c:\Rl \ni t \mapsto \a^t_C(x) \in \cm$ the following holds:
\ba
\frac{d}{dt}\a^t_C(x)=\chi_C(\a^t_C(x))  \q .
\ea
The flow satisfies the group property $\a^s_C\circ\a^t_C=\a^{(s+t)}_C$. It also acts on phase space functions, that is we have a map $a^t_C:\cc^\infty \ni f \mapsto \a^t_C(f) \in \cc^\infty$. The value of the function $\a^t_C(f)$ at the point $x$ is given by
\ba
\a^t_C(f)(x)=f(\a^t_C(x))   \q  . 
\ea 
It can be calculated with the help of the series
\ba
\a^t_C(f)(x)=\sum_{r=0}^\infty \frac{1}{r!}\,\{C,f\}_r (x) 
\ea  
where $\{C,f\}_0:=f$ and $\{C,f\}_{r+1}=\{C,f\}_{r}$.

Now a gauge transformation $\fg:\cm \ni x \mapsto \fg(x)\in \cm$ is a transformation which can be written as a composition of flows generated by the constraints $C_j,\,j=1,\ldots,n$. The first class property (\ref{1pre1}) guarantees that for $x\in \cc$ the set $\{\fg(x) \,|\,\fg \;\text{is a gauge transformation}\}$ is an $n$-dimensional submanifold of the constraint hypersurface $\cc$ -- the gauge orbit through $x$, which we will denote by $\cg_x$. (For constraint algebras with structure constants a gauge orbit can also be defined for points not contained in the constaint hypersurface $\cc$.) 

A physical (classical) state is an equivalence class of phase space points, where the point $x$ is equivalent to $y$ if $y\in \cg_x$. Hence a physical state can be identified with an $n$-dimensional gauge orbit. Moreover physical states are restricted to the $(2p-n)$ dimensional constraint hypersurface $\cc$, which shows that the space of physical states is $(2p-2n)$ dimensional.

From an physical observable $F$ (i.e. a phase space function) one would expect that its value is well defined by the physical state, i.e. that it does not give different values on gauge equivalent phase space point. This translates into the condition
\ba \label{1pre7}
\{C_j,F\}=0 \q \text{for}\; j=1,\ldots n
\ea
which has to be fulfilled on the constraint hypersurface $\cc$. Such functions are called strong Dirac observables, if (\ref{1pre7}) is satisfied everywhere on $\cm$, otherwise weak Dirac observables. On the other hand since admissible physical states are restricted to $\cc$ one can identify functions $f,f'$ which coincide on $\cc$ but differ elsewhere. In such a case we will write $f\simeq f'$ and call $f,f'$ weakly equal. This defines an equivalence relation on the space of phase space functions. Hence the space of physical observables is the space of equivalence classes of Dirac observables. For this space the Poisson bracket is well defined, i.e. does not depend on the representatives of the equivalence classes. 

Finally we would like to mention that one can replace the set of constraints $\{C_1,\ldots,C_n\}$ by another set of constraints $\{\tilde{C}_1,\ldots,\tilde{C}_n\}$ as long as the constraint hypersurfaces defined by these two sets coincide. This is guaranteed if one can write $\tilde{C}_j=\sum_k A_{jk} C_k$ where $(A_{jk})_{j,k=1}^n$ is a matrix of phase space functions with non-vanishing determinant on $\cc$. The two sets of constraints lead also to the same gauge orbits $\cg_x$ if $x \in \cc$. Moreover weak Dirac observables with respect to the first set are also weak Dirac observables with respect to the second and vice versa. This does not hold for strong Dirac observables.

\section{Complete and Partial Observables for Systems with one Constraint}\label{oneconstr}

The concepts of partial observables and complete observables 
were introduced by Rovelli in \cite{RovPartObs}. 
 A partial observable is understood as ``a physical quantity to which we can associate a (measuring) procedure leading to a number'' ,\cite{RovPartObs}. We will assume here, that one can associate to an arbitrary phase space function such a measuring procedure. A partial observable is then a phase space function, which does not need to be a Dirac observable, i.e. it does not have to commute with the constraints. 

A complete observable is ``a quantity whose value can be predicted by the theory (in classical theory)''. We will understand here under a complete observable phase space functions which commute (weakly) with the constraints, i.e. phase space functions, that are invariant under gauge transformations generated by the constraints.

As outlined in \cite{RovPartObs} if one has a system with one constraint $C(x)$, one can associate to two partial observables (that is phase space functions) $f(x),T(x)$ a family of complete observables $F_{[f,T]}(\tau,x)$ labeled by a parameter $\tau$.  This complete observable is defined in the following way: Consider the flow $\alpha_C^t(x)$ generated by the constraint $C$ starting from the phase space point $x$. The function $F_{[f,T]}(\tau,x)$ gives the value that the function $f_x(t):=f(\alpha_C^t(x))$ assumes if the function $T_x(t):=T(\alpha_C^t(x))$ assumes the value $\tau$. Hence the definition is
\be \label{2defequ1}
F_{[f,T]}(\tau,x)=\alpha_C^t\,(f)(x)_{|\alpha_C^t(T)(x)=\tau}
\ee
One can interprete $T$ as a kind of clock, whose values parametrize the gauge orbit of $C$. The complete observable $ 
F_{[f,T]}(\tau,x)$ predicts the value of $f$ for the ``time'' $\tau$. Here we can already see the conditions under which $F$ is well defined:  
The function $T$ has to provide a good parametrization of the gauge orbit through $x$, that is the function $T_x(t):=\alpha_C^t(T)(x)$ has to be invertible. Now, locally this will be the case as long as  
\be
\frac{d}{dt} \alpha_C^t(T)(x) =\alpha_C^t\left(\{C,T\}\right)(x)\neq 0 \q .
\ee
However, $T_x(t)$ does not have to be globally invertible, it suffices that the function $f_x(t):=\alpha_C^t(f)(x)$ fulfils $f_x(t)=f_x(s)$ for all $s,t$ for which $T_x(t)=T_x(s)$.

One can get a still weaker condition by considering a fixed value of $\tau$. Then $f_x(t_i)$ has to be the same for all values $t_i$ in the preimage $(T_x)^{-1}(\tau)$.

We will now prove that $F_{[f,T]}(\tau,x)$ is indeed a Dirac observable. 
To this end we have to show that $\alpha_C^\veps(F_{[f,T]}(\tau,\cdot))=F_{[f,T]}(\tau,\cdot) $:
\ba
\alpha_C^\veps(F_{[f,T]}(\tau,\cdot))(x) = F_{[f,T]}(\tau,\alpha_C^\veps(x))
&=&\alpha_C^t(f)(\alpha^\veps_C(x))_{|\alpha_C^t(T)(\alpha_C^\veps(x))=\tau} \nn \\
&=&\alpha^\veps_C \circ \alpha^t_C (f)(x)_{|\alpha_C^\veps \circ \alpha_C^t(T)(x)=\tau} \nn \\
&=&\alpha^{t+\veps}_C (f)(x)_{|\alpha_C^{t+\veps}(T)(x)=\tau}
\ea
In the last expression, $s=t+\veps$ is just a dummy variable -- one has to solve $\alpha_C^{s}(T)(x)=\tau$ for $s$ and then to replace $s$ in $\alpha^{s}_C (x)$. This term is therefore equal to $F_{[f,T]}(\tau,x)$ and we have proved the theorem:
\begin{Theorem} \label{T1.1}
Let $f,T$ be two phase space functions and $x \in \cm$ a phase space point, fulfilling the condition: $\alpha_C^t(f)(x)= \alpha_C^s(f)(x)$ for all values $s,t \in \Rl$ for which $\alpha_C^t(T)(x)= \alpha_C^s(T)(x)$. Then $F_{[f,T]}(\tau,x)$ is invariant under the flow generated by $C$.
\end{Theorem}
\vspace{0.4cm}
{\bf Example 1:}
We will consider the constraint  
\be
C=q_1 p_2-q_2 p_1
\ee
on phase space $\cm=\Rl^2_q \times \Rl^2_p$. We will choose $f=q_2$ and $T=q_1$ as partial observables. To calculate the associated complete observable, we have to evolve both partial observables under the flow of $C$:
\ba
\alpha_C^t(T)(q_1,q_2,p_1,p_2)&=&q_1 \cos(t)+q_2 \sin(t)=\sqrt{q_1^2+q_2^2}\sin\left(t+\arctan(q_1/q_2)\right) \nn \\
\alpha_C^t(f)(q_1,q_2,p_1,p_2)&=&q_2 \cos(t)-q_1 \sin(t)=\sqrt{q_1^2+q_2^2}\cos\left(t+\arctan(q_1/q_2)\right)  \q .
\ea
Now we have to invert $T_x(t):=\alpha_C^t(T)(x)$ and to find all values in the preimage of $\tau$. Since $T_x(t)$ is a periodic function the inverse will be a multivalued function. The equation $T_x(t)=\tau$ is uniquely solvable on the interval $t\in \left[-\tfrac{\pi}{2}-\arctan(\tfrac{q_1}{q_2}),\tfrac{\pi}{2}-\arctan(\tfrac{q_1}{q_2})\right]$, where the solution is given by
\be
t_{10}=\arcsin(\frac{\tau}{\sqrt{q_1^2+q_2^2}})-\arctan(\frac{q_1}{q_2}) \q .
\ee
(If not otherwise specified, we will always take the positive branch of the square root.) All other solutions are given by 
\ba
t_{1k}&=&t_{10}+2\pi k \nn \\
t_{2k}&=&\pi-t_{10}-2\arctan(\frac{q_1}{q_2})+2\pi k  \q , \q\q k \in \Zl \q .
\ea
Evaluating $f_x(t):=\alpha_C^t(f)(x)$ at these points gives
\ba
f_x(t_{1k})=\sqrt{q_1^2+q_2^2-\tau^2}\q\q\text{and}\q\q f_x(t_{2k})=-\sqrt{q_1^2+q_2^2-\tau^2} \q ,
\ea
hence $F_{[f=q_2,T=q_1]}(\tau,x)=\pm \sqrt{q_1^2+q_2^2-\tau^2}$ is a double valued function. Even thought the condition in Theorem \ref{T1.1} is not exactly fulfilled both branches of $F_{[q_2,q_1]}(\tau,x)$ are Dirac observables, i.e. commute with the constraint $C$.
\vspace{0.4cm}

This example suggests to introduce a two-dimensional configuration space of partial observables $\cn$, which is coordinatized by values of the partial observables $T$ and $f$.\footnote{The topology of $\cn$ should be determined by the properties of the partial observables.} That is, we have a map $P:\cm\rightarrow\cn$ given by $x\mapsto(T(x),f(x))$. Fix a point $x\in \cm$. Then the flow $\alpha_C^t(x)$ of the point $x$ in $\cm$ induces a flow of the point $P(x)=(T(x),f(x))$ by $\alpha_C^t(T(x),f(x)):=P(\alpha_C^t(x))=(T(\alpha_C^t(x)),f(\alpha_C^t(x)))$. One has to keep in mind that this flow in $\cn$ is not necessarily uniquely determined by the initial point $(T(x),f(x))\in \cn$ but it is determined by the initial point in $x \in \cm$ (see next example).

In this way each point in $\cm$ defines a gauge orbit in $\cn$, namely the set $\{P(\alpha_C^t(x))\,|t\in \Rl\}$. (Of course, gauge equivalent points in $\cm$ define the same gauge orbit in $\cn$. It may also happen, that gauge inequivalent points in $\cm$ define the same gauge orbit in $\cn$ as is the case in example 1, where the gauge orbits do not depend on the momenta.) The functions $(T_x(t)=T(\alpha_C^t(x)),f_x(t)=f(\alpha_C^t(X))$ provide a parameter description of this gauge orbit. One can interpret the complete observable $F_{[f,T]}(x,\cdot)$ as a function of the first coordinate of $\cn$, whose graph $(\cdot,F_{[f,T]}(x,\cdot))$ coincides with the gauge orbit. Of course, it is in general not possible to describe a curve (namely the gauge orbit) by a graph of a single valued function -- exactly this is the case in example 1, where one needs a double valued function. Another way to describe a surface or curve is as a level set of a function from $\cn$ to $\Rl$. In example 1 this desciption is given by $T(x)^2+f(x)^2=q_1^2+q_2^2=\text{const.}$.    

\vspace{0.4cm}
{\bf Example 2:}
Here we will consider the constraint
\be
C=\tfrac{1}{2}(p_1^2+\omega_1 q_1^2)-\tfrac{1}{2}(p_2^2+\omega_2 q_2^2) 
\ee
where $\omega_1,\omega_2$ are (constant) frequencies. The phase space is again $\cm=\Rl^2_q\times \Rl^2_p$. The clock variable $T$ will be $T(x)=q_1$, and the other partial observable is $f(x)=q_2$. The evolution of these partial observables under the flow of $C$ is again periodic: 
\ba
\alpha_C^t(T)(q_1,q_2,p_1,p_2)&=&q_1 \cos(\omega_1 t)-p_1 \sin(\omega_1 t)=\sqrt{q_1^2+(p_1/\omega_1)^2}\sin\left(\omega_1 t-\arctan(\omega_1q_1/p_1)\right) \nn \\
\alpha_C^t(f)(q_1,q_2,p_1,p_2)&=&q_2 \cos(\omega_2 t)-p_2 \sin(\omega_2 t)=\sqrt{q_2^2+(p_2/\omega_2)^2}\sin\left(\omega_2 t+\arctan(\omega_2q_2/p_2)\right)  \; . \nn \\
\ea 
The function $T_x(t)=\alpha_C^t(T)(x)$ is uniquely invertible on the intervall $[-\tfrac{\pi}{2\omega_1}+\arctan(\omega_1 q_1/p_1), \tfrac{\pi}{2\omega_1}+\arctan(\omega_1 q_1/p_1)]$ (and all intervals which one can get by translating the first interval by an amount $k\pi, k \in \Zl$). The solutions of $T_x(t)=\tau$ are given by 
\ba
t_{1k}&=&\frac{1}{\omega_1}\left(\arcsin\left(\frac{\omega_1 \tau}{\sqrt{\omega_1^2q_1^2+p_1^2}}\right)+\arctan\left(\frac{\omega_1 q_1}{p_1}\right)+2\pi k\right) \nn \\
t_{2k}&=&\frac{1}{\omega_1}\left(\pi-\arcsin\left(\frac{\omega_1 \tau}{\sqrt{\omega_1^2q_1^2+p_1^2}}\right)+\arctan\left(\frac{\omega_1 q_1}{p_1}\right)+2\pi k\right) \q\q k \in \Zl \q .
\ea
Applying the function $f_x(\cdot)=\alpha_C^t(f)(x)$ to these values we will get the complete observable $F_{[f,T]}(\tau,x)$, which in general will be multi-valued. We will label these multiple values by $\{i,k\},i=1,2;k\in \Zl$:
\ba \label{E1.2result}
F_{[f,T]}(\tau,x)_{1k}&=&\sqrt{q_2^2+(\frac{p_2}{\omega_2})^2}\,\sin\Bigg( \frac{\omega_2}{\omega_1}\left( \arcsin\left(\frac{\omega_1 \tau}{\sqrt{\omega_1^2q_1^2+p_1^2}}\right)+\arctan\left(\frac{\omega_1 q_1}{p_1}\right)+2\pi k\right) \nn \\
&& \q\q\q\q\q\q\q\q\q+\arctan\left(\frac{\omega_2 q_2}{p_2}\right) \Bigg) 
\nn \\
F_{[f,T]}(\tau,x)_{2k}&=&\sqrt{q_2^2+(\frac{p_2}{\omega_2})^2}\,\sin\Bigg(\frac{\omega_2}{\omega_1}
\left( \pi -
\arcsin\left(\frac{\omega_1 \tau}{\sqrt{\omega_1^2q_1^2+p_1^2}}\right)+\arctan\left(\frac{\omega_1 q_1}{p_1}\right)+2\pi k\right) \nn \\
&& \q\q\q\q\q\q\q\q\q+\arctan\left(\frac{\omega_2 q_2}{p_2}\right) \Bigg) \q .
\ea
In spite of the multi-valuedness of $F_{[f,T]}(\tau,x)$ all the phase space functions $F_{[f,T]}(\tau,x)_{i,k}$ commute with the constraint $C$. (All the functions in (\ref{E1.2result}) are combinations of $h_j(x):=q_j^2+(p_j/\omega_j)^2,j=1,2$ and 
\be
g(x):=\frac{\omega_2}{\omega_1}  \arctan\left(\frac{\omega_1 q_1}{p_1}\right)+\arctan\left(\frac{\omega_2 q_2}{p_2}\right) 
\ee
and $C$ commutes with $h_j$ and $g$.)

In the projected phase space $\cn=\Rl^2$ the gauge orbits of $C$ are Lissajous figures. If the frequencies $\omega_1$ and $\omega_2$ are non-commensurable these curves fill densely the rectangle\\ $[-\sqrt{q_1^2+(p_1/\omega_1)^2},\sqrt{q_1^2+(p_1/\omega_1)^2}]\times [-\sqrt{q_2^2+(p_2/\omega_j)^2},\sqrt{q_2^2+(p_2/\omega_2)^2}]$.\\
Therefore it is not astonishing that the prediction of $f_x(t)$ given the value of $T_x(t)$ is highly ambiguous. Nevertheless it is possible to obtain in this way a Dirac observable.
\vspace{0.4cm}

In practice it will not be possible to obtain such global information on the behavior of the partial observables $T,f$ as is needed for the proof of the theorem \ref{T1.1}. In the above example we could have done the following: To invert the flow of the clock variable only on some suitable small interval and to calculate the complete observable just for parameter values in this interval. We would have obtained not all values of the multi-valued complete observable but just one of them. Nevertheless in the above example this value turns out to be a Dirac observable.

The questions arises whether this argumentation works also in more complicated or chaotic system. Of course one could take the viewpoint that one can at least try to calculate complete observables and then check afterwards whether the complete observables obtained are Dirac observables, i.e. commute with the constraints. We will proceed in this way in some examples for practical reasons. The next example gives a glimpse of what might happen in chaotic systems.  
\vspace{0.4cm}

{\bf Example 3:}
The Partial Observable Method promises to give a Dirac observable (or constant of motion if one identifies the constraint with the Hamiltonian) for arbitrary pairs of phase space functions satisfying the conditions mentioned in theorem \ref{T1.1}. But we know that in ergodic systems there do not exist analytic constants of motions (see \cite{chaos}). It would be therefore interesting to know how the method applies to such systems. Since it is very difficult to perform explicit calculations for such systems, we will leave this question open for further research and consider here a very simple ergodic systems, which evolves in discrete steps.

We will investigate the so called Baker transformation, see for instance \cite{ReedSimon}. This system is defined on the phase space $\cm=[0,1]_{x_1}\times[0,1]_{x_2}$ (identified to a torus) and evolves in discrete steps in the following way
\ba
\alpha_1((x_1,x_2))=
\begin{cases}
 &(2x_1,\tfrac{1}{2}x_2)  \q\q\q\q\q  \text{if}  \q  0 \leq x_1 <\tfrac{1}{2}\\
 &(2x_1-1,\tfrac{1}{2}x_2+\tfrac{1}{2})      \q\;  \text{if} \q \tfrac{1}{2} \leq x_1 <1       
\end{cases}
\ea
where $\a_K(\cdot)$ denotes the evolution after $K$ steps.

As partial observables we choose $T=x_1$ and $f=x_2$, i.e the phase space coordinates. Now one has to consider the flow of these functions. 

Assume that $x_1 <2^{-N}$ for some number $N\in \Nl$. Then $\a_K(T)(x)=2^K x_1$ for $K\leq N$ and we can invert the function $T_x(K):=\a_K(T)(x)$ on the interval $K\in [0,N]$: The solution of the equation $T_x(K)=\tau$  in this interval is
\be
K_0=\log_2 (\tau/x_1) \q .
\ee
Inserting this solution into the function $f_x(K):=\a_K(f)(x)=2^{-K}x_2$ gives
\be
F_{[f,T]}(\tau,x)=f_x(K_0)= \frac{x_1\cdot x_2}{\tau} \q .
\ee
Now, assuming $x_1 <2^{-N}$, $F_{[f,T]}(\tau,x)$ is not invariant under the flow $\alpha_J(\cdot)$ for arbitrary $J \in \Nl$, but it is invariant for $J\leq N$. Hence these $J$'s lie exactly in the interval, on which we inverted the function $T_x(K)$.  

Beginning with the assumption $1>x_1 \geq 1-2^{-N}$ one can invert the function $T_x(K)$ on the interval $K\in[0,N]$ and one finds the complete observable
\be
F'_{[f,T]}(\tau,x)=1-\frac{(1-x)(1-y)}{1-\tau}  \q .
\ee
Again, assuming $x_1 \geq 1-2^{-N}$, this phase space function is invariant under the flow $\alpha_J(\cdot)$ for $J\in[0,N]$. 

The complete observable $F_{[f,T]}(\tau,x)$ for $x_1 <2^{-N}$ corresponds to the fact that the phase space function $g(x)=x_1\cdot x_2$ is invariant under the evolution law which applies in the region $x_1<\tfrac{1}{2}$, whereas $F'{[f,T]}(\tau,x)$ corresponds to the phase space function $g'(x)=(1-x_1)(1-x_2)$ which is invariant under the evolution law which applies in the region $x_1\geq\tfrac{1}{2}$. \;

\vspace{0.4cm}

Because the above model evolves in discrete steps it is difficult to draw conclusions for continuous models. It is possible to show that for certain models there do not exist analytic single-valued constants of motion, see \cite{chaos}. So if one would calculate complete observables for those models the result may be still a Dirac observable, but a non-analytic and multi-valued one.

\section{Complete and Partial Observables for Systems with Several Constraints}
\label{manyconstr}

The aim of this chapter is to define complete observables for systems with several constraints. For systems with one constraint we had the following geometrical picture: The constraint $C$ generates a one-dimensional gauge orbit. We used the ``clock'' variable $T$ to parametrize this gauge orbit. In the ideal case each point of the gauge orbit was uniquely specified by a value $\tau$ of the phase space function $T$. We then evaluated the function $f(\cdot)$ at the phase space point labelled by $\tau$.

Now for a system with $n$ independent (first class) constraints the gauge orbits generated by these constraints will be $n$-dimensional.\footnote{The first class property of the constraints guarantees, that the flow of the constraints is integrable to an $n$ dimensional surface -- the gauge orbit. For constraint algebras with structure functions the integrability conditions are only fulfilled on the constraint hypersurface, hence one can only expect on the constraint hypersurface well defined gauge orbits. In this case the method following below will work only for points $x$ on the constraint hypersurface. Moreover if one replaces the constraints by another equivalent set of constraints, as was mentioned in section \ref{prelim}, the new constraints will only generate the same gauge orbits as the old constraints on the constraint hypersurface.}
  Hence we need $n$ clock variables (i.e. phase space functions) $T_j,j=1,\ldots,n$ to parametrize a gauge orbit. Ideally each point of a (fixed) gauge orbit should be uniquely specified by the values $(\tau_1,\ldots,\tau_n)$ of the clock variables $(T_1,\ldots,T_n)$. It is then non-ambiguous to evaluate a phase space function $f$ at that point of a (fixed) gauge orbit, on which the functions $(T_1,\ldots,T_n)$ give the values $(\tau_1,\ldots,\tau_n)$. Locally this amounts to the condition
\be
\text{det}(\{C_j,T_i\}_{i,j=1}^n)\neq 0 
\ee
which has to be satisfied on the gauge orbit through $x$.

Again, the above procedure is still well defined, if to the values $(\tau_1,\ldots,\tau_n)$ correspond several points of the gauge orbit but if $f$ evaluated at these points gives always the same number. Otherwise the result will be (possibly extremely) multi-valued. 

Analogously to the previous section we therefore define the complete observable $F_{[f,T_1,\ldots,T_n]}$ associated to the partial observables $f,T_1,\ldots,T_n$ as
\be \label{2Def1}
F_{[f,T_1,\ldots,T_n]}(\tau_1,\ldots,\tau_n,x)=\alpha_{\beta_j C_j}(f)(x)_{|\alpha_{\beta_j C_j}(T_i)(x)=\tau_i} \q .
\ee
Here $\alpha_{\beta_j C_j}(\cdot)$ is defined to be $\a_{\sum_j\b_j C_j}^{t=1}(\cdot)$, that is the flow generated by the phase space function $\sum_j \b_jC_j$ evaluated at the parameter value $t=1$. 
Here and in the following we will assume that all points of the gauge orbit $\cg_x$ can be reached by transformations of the form $\{\a_{\b_jC_j}(x);\,\b_j\in\Rl\}$. (We defined $\cg_x$ to be the set which can be reached by all forms of compositions of flows of the form $\a_{\b_jC_j}(x)$.)

The complete observable $F_{[f,T_1,\ldots,T_n]}(\tau_1,\ldots,\tau_n,x) $ is not defined for those parameter values $\tau_i$ for which the intersection of the level set $\{x\;|T_i(x)=\tau_i \,\, \text{for}\,\, i=1,\ldots,n\}$ and the gauge orbit through $x$ is empty.

To calculate the complete observable $F_{[f,T_1,\ldots,T_n]}$ one has at first to solve the system of equations 
\be \label{2Def1a}
\alpha_{\b_jC_j}(T_i )(x)=\tau_i  \q, i=1,\ldots,n
\ee
for $\b_1,\ldots,\b_n$. Next one has to plug in these solutions $\b_k=B_k(\tau_i,x)$ 
into the function $f_x(\b_1,\ldots,\b_n):=\alpha_{\beta_j C_j}(f)(x)$. This gives the complete observable  $F_{[f,T_1,\ldots,T_n]}(\tau_1,\ldots,\tau_n,x)=f_x(B_1(\tau_i,x),\ldots,B_n(\tau_i,x))$. 

Different solutions $\b_k=B_k(\tau_i,x)$ of (\ref{2Def1a}) may correspond to the same phase space point on the gauge orbit in question or to different points of this gauge orbit. In the latter case $F_{[f,T_1,\ldots,T_n]}(\tau_i,x)$ is still well defined, if $f(\cdot)$ evaluated at these points gives always the same number. Otherwise $F_{[f,T_1,\ldots,T_n]}(\tau_i,x)$ will be multi-valued.

We will now verify that $F_{[f,T_1,\ldots,T_n]}(\tau_i,x)$ is indeed an $n$-parameter family of Dirac Observables. To this end we have to show that $F$ remains constant under the flow of $\alpha_{\veps_j C_j}(\cdot)=\a_{\sum_j\veps_j C_j}^{t=1}(\cdot)$, where $\veps_j,j=1,\ldots,n$ are arbitrary fixed numbers:
\ba \label{2proof1}
\a_{\veps_j C_j}(F_{[f,T_k]}(\tau_m,\cdot))(x)=F_{[f,T_k]}(\tau_m,\a_{\veps_j C_j}(x) )&=&\a_{\b_jC_j}(f)(\a_{\veps_lC_l}(x))_{|\a_{\b_jC_j}(T_i)(\a_{\veps_lC_l}(x))=\tau_i} \nn \\
&=&\a_{\veps_l C_l}\circ\a_{\b_j C_j}(f)(x)_{|\a_{\veps_l C_l}\circ\a_{\b_j C_j}(T_i)(x)=\tau_i} \; .\q\q
\ea
The key point in the last expression is that $\a_{\veps_l C_l}\circ\a_{\b_j C_j}(x)$ is still a point on the gauge orbit $\cg_x$ through $x$. (For constraint algebras with structure functions this holds only on the constraint hypersurface.) Hence solutions of $\a_{\veps_l C_l}\circ\a_{\b_j C_j}(T_i)(x)=\tau_i$ correspond to the same points on the gauge orbit 
as solutions of $\a_{\b_j C_j}(T_i)(x)=\tau_i$  (namely to the intersection of the level sets $\{T_i^{-1}(\tau_i)\}_{i=1}^n$ and $\cg_x$).
Both expressions in (\ref{2proof1}) and (\ref{2Def1a}) prescribe to evaluate the phase space function $f(\cdot)$ at exactly these points. If $f(\cdot)$ gives always the same value at these points the expressions  in (\ref{2proof1}) and (\ref{2Def1a}) are well defined and coincide. Thus we have the theorem:
\begin{Theorem}   \label{theorem2}    
Let $C_j,j=1,\ldots,n$ be $n$ independent first class constraint and $x \in \cm$ a phase space point on the constraint hypersurface. Given $n$ phase space functions $T_i,i=1,\ldots,n$ and a phase space function $f$, assume that $f$ evaluated at the points in the intersection of the level set $\{x\;|T_i(x)=\t_i\,\,\text{for} \,\, i=1,\ldots,n\}$ and the gauge orbit through $x$ gives always the same result. Then the complete observable $F_{[f,T_i]}(\tau_i,x)$ is well defined and is invariant under the flow generated by the constraints.
\end{Theorem}

Analogously to the last section we can introduce a configuration space of partial observables, which is an $(n+1)$-dimensional manifold coordinatized by values of the partial observables $(T_1,\ldots,T_n,f)$. The ``projection'' $P:\cm \rightarrow\cn$ is defined by $P: x \mapsto (T_1(x),\ldots,T_n(x),f(x))$. 

Fix a point $x$. Then the flow $\a_{\b_jC_j}(x)$ induces a flow of the point $(T_1(x),\ldots,T_n(x),f(x))$ in $\cn$ by $\a_{\b_jC_j}((T_1(x),\ldots,T_n(x),f(x)))=P(\a_{\b_jC_j}(x))$. Again one has to keep in mind that this flow needs not to be uniquely determined by the initial point $(T_1(x),\ldots,T_n(x),f(x)) \in \cn$, but it is uniquely determined by the point $x \in \cm$. Since in $\cm$ the flow generated by the constraints integrates to a hypersurface, namely the gauge orbit (at least on the constraint hypersurface), this is also the case in the space $\cn$. This gauge orbit can be described in a parametric way, i.e. as the set $\{P(\a_{\b_jC_j}(x))\,|\b_j \in \Rl\}$. The complete observable $F_{[f,T_i]}(\tau_i,x)$ provides a description of the gauge orbit as the graph of a function depending on $n$ variables, i.e. the gauge orbit is the set 
$\{(\tau_1,\ldots,\tau_n,F_{[f,T_i]}(\tau_1,\ldots,\tau_n,x))\,|\tau_1,\ldots,\tau_n\in \Rl\}$.

\vspace{0.4cm}

{\bf Example 4:} 
Here we will consider the phase space $\cm=\Rl_q^3\times\Rl_p^3$ and take the coordinates of the angular momentum as constraints:
\be \label{E3.1con}
C_i=\sum_{j,k=1}^3 \eps_{ijk}\,q_jp_k \q ,i=1,2,3
\ee
where $\eps_{ijk}$ is the totally antisymmetric tensor with $\eps_{123}=1$. 

But the three constraints (\ref{E3.1con}) are not algebraically independent since $\sum x_i C_i=\sum p_i C_i=0$. Hence it is sufficient to consider $C_1$ and $C_2$, which form a constraint algebra with structure functions:
\be
\{C_1,C_2\}=C_3=-\frac{1}{q_3}(q_1C_1+q_2C_2)=-\frac{1}{p_3}(p_1 C_1+p_2 C_2) \q .
\ee
Therefore we need two clock variables which we will choose as $T_1=q_1$ and $T_2=q_2$. The flow of these variables generated by the constraints is
\ba
\a_{(\b_1C_1+\b_2C_2)}(T_1)(x)\!\!&=&\!\!q_1+\frac{\b_2(\b_2q_1-\b_1q_2)}{\b_1^2+\b_2^2}\left(\!\cos(\sqrt{\b_1^2+\b_2^2})-1\!\right)-\frac{\b_2q_3}{\sqrt{\b_1^2+\b_2^2}}\sin(\sqrt{\b_1^2+\b_2^2}) \nn \\
\a_{(\b_1C_1+\b_2C_2)}(T_2)(x)\!\!&=&\!\!q_2+\frac{\b_1(\b_1q_2-\b_2q_1)}{\b_1^2+\b_2^2}\left(\!\cos(\sqrt{\b_1^2+\b_2^2})-1\!\right)+\frac{\b_1q_3}{\sqrt{\b_1^2+\b_2^2}}\sin(\sqrt{\b_1^2+\b_2^2}) \nn \\
\ea
Next we have to solve the system of equations
\ba \label{E3.1diffi}
\a_{(\b_1C_1+\b_2C_2)}(T_1)(x)&=&\tau_1 \nn \\
\a_{(\b_1C_1+\b_2C_2)}(T_2)(x)&=&\tau_2    \q 
\ea
for $\b_1$ and $\b_2$. But already in this relatively simple example, it is very difficult to find the solutions of (\ref{E3.1diffi}), since on the left hand side the $\b_i$'s appear in the argument and as coefficients of transcedental functions. In the next section we will explain a different method to calculate complete observables, but here we will use that one can replace the constraints $C_1$ and $C_2$ by an equivalent set of constraints, obtained from the first ones by multiplying them with (nowhere-vanishing) phase space functions. 

In this example we will use the modified constraints
\ba
C'_1=\frac{1}{q_2}C_1=p_3-\frac{q_3 p_2}{q_2} \q\q\q C'_2=-\frac{1}{q_1}C_1=p_3-\frac{q_3 p_1}{q_1}   \q .
\ea
The flow generated by these constraints is especially simple for the phase space function $q_3$, so this will be one of our clock variables $T_1:=q_3$:
\ba
\a_{\b_1C'_1+\b_2C'_2}(T_1)(x)=q_3-\b_1-\b_2 \q .
\ea
To find a suitable second clock observable, we will consider the flow of the phase space function $g=q_1$. Its Poisson bracket with the constraints is
\be
\{\b_1 C'_1+\b_2 C'_2, q_1\}= \b_2 \frac{q_3}{q_1} \q .
\ee
The function $q_1$ in the denominator of the right hand side vanishes if we choose the function $T_2:=g^2=q_1^2$ as a second clock variable. One obtains the following differential equation for the flow $T_{2x}(t):=\a_{\b_1 C'_1+\b_2 C_2}^t(T_2)(x)$:
\be
\frac{d}{dt}T_{2x}(t)=2\b_2 \, T_{1x}(t)
\ee
where $T_{1x}(t):=\a_{\b_1 C'_1+\b_2 C_2}^t(T_1)(x)=q_3-(\b_1+\b_2)t$. This differential equation is easily integrable and gives for the flow of the second clock variable

\be
\a_{\b_1C'_1+\b_2C'_2}(T_2)(x)=T_{2x}(t=1)=q_1^2-\b_2(\b_1+\b_2)+2\b_2q_3 \q .
\ee
Analogously, choosing as the third partial observable $f=q_2^2$, its evolution under the constraints is
\be \label{E3.1evolf}
\a_{\b_1C'_1+\b_2C'_2}(f)(x)=q_2^2-\b_1(\b_1+\b_2)+2\b_1q_3 \q .
\ee
Now we have to solve the system of equations
\begin{alignat}{2}
\a_{\b_1C'_1+\b_2C'_2}(T_1)(x)\,&=\,q_3-\b_1-\b_2&=\tau_1 \nn \\
\a_{\b_1C'_1+\b_2C'_2}(T_2)(x)\,&=\,q_1^2-\b_2(\b_1+\b_2)+2\b_2q_3\,&=\,\tau_2
\end{alignat}
for the parameters $\b_1$ and $\b_2$. The solutions are given by
\ba
\b_1=B_1(\tau_1,\tau_2,x)&=&\frac{q_1^2+q_3^2-\tau_1^2-\tau_2}{q_3+\tau_1} \nn \\
\b_2=B_2(\tau_1,\tau_2,x)&=&\frac{\tau_2-q_1^2}{q_3+\tau_1} \q \q\q\q\q.
\ea
Replacing the $\b_i$'s in the evolution of $f$ in (\ref{E3.1evolf}) by the solutions $B_i(\tau_1,\tau_2,x)$ we obtain finally the complete observable
\be
F_{[f,T_1,T_2]}(\tau_1,\tau_2,x)=q_1^2+q_2^2+q_3^2-\tau_1^2-\tau_2 \q .
\ee
This phase space function is indeed a Dirac observable.

\section{A System of Partial Differential Equations for Complete Observables}
\label{PDE}

As we have seen in example 4 it may be very difficult to invert the flow of the clock variables $T_i$ in order to find the complete observable $F_{[f,T_i]}$. In this section we will derive a system of (first order) partial differential equations for $F_{[f,T_i]}(\tau_j,x)$ as a function of the $\tau_j$'s.\footnote{For systems with one constraint such a differential equation for a complete observable appeared in \cite{problemoftime}.} To this end we will make the following assumptions: Consider a phase space point $x\in \cm$ and the gauge orbit $\cg_x$ through $x$ in $\cm$. We define the map 
\ba
{\bf T}_x:\cg_x &\rightarrow& {\bf T}_x(\cg_x) \subset \Rl^n \nn \\
             y  &\mapsto & (T_1(y),\ldots,T_n(x))   \q .
\ea
The index $x$ in ${\bf T}_x$ represents the gauge orbit $\cg_x$ through $x$. Our assumption is, that ${\bf T}_x$ is uniquely invertible as a function from $\cg_x$ to ${\bf T}_x(\cg_x)$. In other words, to each point $(\tau_1,\ldots \tau_n)\in{\bf T}_x(\cg_x)$ there exists a unique point $y \in \cg_x$ which solves $T_k(y)=\tau_k$ for $k=1,\ldots,n$. We will denote this point by $y={\bf T}_x^{-1}(\tau_i)$. 

We defined the complete observable $F_{[f,T_i]}$ as
\ba
F_{[f,T_i]}(\tau_i,x)=\a_{\b_k C_k}(f)(x)_{|\a_{\b_k C_k}(T_i)(x)=\tau_i} \q .
\ea
The value of $F_{[f,T_i]}$ at the slightly displaced point $(\tau_1+\veps_1,\ldots,\tau_n+\veps_n)$ is
\ba \label{4pde1}
F_{[f,T_i]}(\tau_i+\veps_i,x)=\a_{\b_k C_k}(f)(x)_{|\a_{\b_k C_k}(T_i)(x)=\tau_i+\veps_i }\q .
\ea
Now we know that $F_{[f,T_i]}$ is gauge invariant, therefore we can replace $x$ by the point ${\bf T}_x^{-1}(\tau_i)\in \cg_x$ on the right hand side of (\ref{4pde1}):
\ba
F_{[f,T_i]}(\tau_i+\veps_i,x)=\a_{\b_k C_k}(f)({\bf T}_x^{-1}(\tau_i) )_{|\a_{\b_k C_k}(T_j)({\bf T}_x^{-1}(\tau_i) )=\tau_j+\veps_j }\q
\ea
As by definition $T_j({\bf T}_x^{-1}(\tau_i) )=\tau_j$, we can solve the equations
\ba\label{4pde5}
\a_{\b_k C_k}(T_j)({\bf T}_x^{-1}(\tau_i) )=\tau_j+\sum_k \b_k \{C_k,T_j\}({\bf T}_x^{-1}(\tau_i) )+\co(\veps^2) =\tau_j+\veps_j 
\ea
for the $\b_k$'s to the first order in the $\veps_i$'s. (Here $\co(\veps^2)$ denotes terms of higher than first order in the $\veps_i$'s.) To this end we define the matrix of phase space functions
\ba \label{4pde6}
A_{kj}:= \{C_k,T_j\} 
\ea
and its inverse $A^{-1}_{jm}$ by $A_{kj}A^{-1}_{jm}=\delta_{km}=A^{-1}_{kj}A_{jm}$. (The inverse exists at least on the gauge orbit $\cg_x$ because of the assumptions we made for the map ${\bf T}_x$.) The solution to the equations (\ref{4pde5}) can then be written as
\ba
\beta_k=\sum_j \veps_j \, A^{-1}_{jk}({\bf T}_x^{-1}(\tau_i) )\; + \co(\veps^2) \q .
\ea
Inserting these values into 
\ba
\a_{\b_k C_k}(f)({\bf T}_x^{-1}(\tau_i) )=f({\bf T}_x^{-1}(\tau_i))+\sum_k \b_k \,\{C_k,f\}({\bf T}_x^{-1}(\tau_i) )   \;+ \co(\b^2)
\ea
we arrive at
\ba
F_{[f,T_i]}(\tau_i+\veps_i,x)&=&
f({\bf T}_x^{-1}(\tau_i))+\sum_{j,k}\veps_j \left(A^{-1}_{jk}\{C_k,f\}\right)({\bf T}_x^{-1}(\tau_i))\;+ \co(\veps^2) \nn \\
&=& F_{[f,T_i]}(\tau_i,x)+\sum_{j,k}\veps_j \left(A^{-1}_{jk}\{C_k,f\}\right)({\bf T}_x^{-1}(\tau_i))\;+ \co(\veps^2)   \q .
\ea
This gives for the partial derivative of $F_{[f,T_i]}$ with respect to $\tau_m$
\ba \label{4pde10}
\frac{\partial}{\partial \tau_m}F_{[f,T_i]}(\tau_i,x)=\left(\sum_{k}   A^{-1}_{mk}\{C_k,f\}\right)({\bf T}_x^{-1}(\tau_i))=:g_m({\bf T}_x^{-1}(\tau_i))  \q .
\ea

An alternative derivation of this result proceeds in the following way: By definition of the complete observable $F_{[f,T_i]}$ we have the equation
\ba
F_{[f,T_i]}(\tau_i=T_i(y),x)=f(y)
\ea  
where $y$ is a point in the gauge orbit $\cg_x$ through $x$. Hence we can also write
\ba 
F_{[f,T_i]}(T_i(\a^t_{\gamma_kC_k}(y)),x)=f(\a^t_{\gamma_kC_k}(y)) \q .
\ea
Differentiating both sides of this equation in $t$ gives
\ba
\sum_m \frac{\partial}{\partial \tau_m} F_{[f,T_i]}(T_i(\a^t_{\gamma_kC_k}(y)),x)  \, \frac{d T_m}{dt}(\a^t_{\gamma_kC_k}(y)  )=\frac{df}{dt}(\a^t_{\gamma_kC_k}(y) ) \q .
\ea
Now we use the defining differntial equation for the flow of a phase space function and set $t=0$:
\ba
\sum_{m,k} \frac{\partial}{\partial \tau_m} F_{[f,T_i]}(T_i(y),x) \, \gamma_k\{C_k,T_m\}(y)=\sum_k \gamma_k\{C_k,f\}(y)  \q .
\ea
This equation has to hold for an arbitrary set of $\gamma_k \in \Rl;\,k=1,\ldots,n$. Hence we can conclude that
\ba
\frac{\partial}{\partial \tau_m} F_{[f,T_i]}(T_i(y),x) \,\{C_k,T_m\}(y)=\{C_k,f\}(y)   \q 
\ea
has to hold for $k=1,\ldots,n$. Multiplying both sides of the equation with the matrix $A^{-1}_{jk}$ (see equation (\ref{4pde6})) and replacing $y$ with $y={\bf T}_x^{-1}(\tau_i)$ we arrive again at equation (\ref{4pde10}): 
\ba
\frac{\partial}{\partial \tau_m}F_{[f,T_i]}(\tau_i,x)=\left(\sum_{k}   A^{-1}_{mk}\{C_k,f\}\right)({\bf T}_x^{-1}(\tau_i))=:g_m({\bf T}_x^{-1}(\tau_i))  \q .
\ea

The function $g_m({\bf T}_x^{-1}(\tau_i))$ on the right hand side of (\ref{4pde10}) can again be written as a complete observable associated to the partial observable $g_m(x)=\sum_{k}   A^{-1}_{mk}\{C_k,f\} $ (and $T_1(x),\ldots,T_n(x)$):
\ba
g_m({\bf T}_x^{-1}(\tau_i))=g_m(\a_{\b_kC_k}(x))_{|T_i(\a_{\b_kC_k}(x))=\tau_i}=F_{[g,T_i]}(\tau_i,x) \q .
\ea
Therefore the partial derivatives with respect to the parameters $\tau_m$ of complete observables are again complete observables (which is not surprising, since a complete observable is a Dirac observable for all values of the parameters $\tau_m$). Now the problem is, that in general the complete observables $F_{[g_m,T_i]}(\tau_i,x)$ are unknown functions. In this case one has to add the partial differential equations (PDE's) for the $F_{[g_m,T_i]}$. Again, it may happen, that these involve unknown functions. In this case one has to iterate the procedure until one obtains a closed system of PDE's for a set of functions $F_{[f,T_i]},F_{[g_m,T_i]},F_{[g_{mm'},T_i]},\ldots$. Using this system it may be possible to derive a higher order PDE for the primary function $F_{[f,T_i]}$.

However one can limit the number of necessary iterations if one realizes that if a function $g$ is composed from $m$ phase space functions $f_h$ the associated complete observable is 
\ba \label{pdecomposed}
F_{[g(f_1,\ldots,f_m), T_i]}(\tau_i,x)&=&\a_{\b_jC_j}(g(f_1,\ldots,f_m))(x)_{|\a_{\b_jC_j}(T_i)(x)=\tau_i}
\nn \\
&=& g(\a_{\b_jC_j}(f_1),\ldots,\a_{\b_jC_j}(f_m))(x)_{|\a_{\b_jC_j}(T_i)(x)=\tau_i} \nn \\
&=&g(F_{[f_1,T_i]}(\tau_i,x),\ldots,F_{[f_m,T_i]}(\tau_i,x))   \q .
\ea
(That is ${\bf{F}}_{[T_i]}(\tau_i): f \mapsto F_{[f,T_i]}(\tau,x)$ is an algebra homomorphism with respect to multiplication and addition.) 

Hence we just need to choose $\text{dim}(\cm)=2p$ algebraically independent phase space functions $f_h$ (for instance the canonical coordinates) and to consider the PDE's (\ref{4pde10}) for the associated complete observables. The right hand side of these PDE's will then be expressible through the complete observables associated to the $f_h,\,h=1,\ldots,2p$. Therefore putting the PDE's for the complete observables associated to the $f_h$'s together one obtains a closed system of partial differential equations for $2p$ unknown functions. Moreover if $n$ of the functions $f_h$ coincide with the constraints and another set of $n$ functions coincides with the clock variables $T_i$, the associated complete observables vanish weakly or are constants respectively:
\ba
F_{[C_k,T_1,\ldots,T_n]}(\tau,x) &\simeq&  0
\nn \\
F_{[T_k,T_1,\ldots,T_n]}(\tau,x)& =&  \tau_k  \q   .
\ea
and one is left with a set of $(2p-2n)$ unknown functions.

The property (\ref{pdecomposed}) shows also that the set
\be
 \{F_{[q_a,T_1,\ldots,T_n]}(\t_1,\ldots,\t_n,\cdot),F_{[p_a,T_1,\ldots,T_n]}(\t_1,\ldots,\t_n,\cdot)\,|\,a=1,\ldots,p\}
\ee
 provides an over-complete basis of the space of Dirac observables: If $d$ is a Dirac observable then
\ba \label{4pdeDirac}
F_{[d,T_i]}(\t_i,x)\simeq d(x=(q_a,p_a))\simeq d(F_{[q_a,T_i]}(\t_i,x),F_{[p_a,T_i]}(\t_i,x))  \q .
\ea
Hence $d$ can be expressed as a combination of the complete observables associated to the canonical coordinates. The above mentioned over-completeness is described by
\ba
F_{[C_k,T_i]}(\t_i,x) &\simeq& 0\simeq C_k(F_{[q_a,T_i]}(\t_i,x),F_{[p_a,T_i]}(\t_i,x)) \nn \\
F_{[T_k,T_i]}(\t_i,x) &\simeq& \t_k\simeq T_k(F_{[q_a,T_i]}(\t_i,x),F_{[p_a,T_i]}(\t_i,x)) \q ,
\ea
that is we have $2n$ relations between the $2p$ complete observables.

 At the end of this section we will give a (formal) solution to the PDE's (\ref{4pde10}) as a power series in the $\tau_i$'s.

Before we come to an example, we will prove, that the PDE's (\ref{4pde10}) are consistent, i.e. satisfy the integrability conditions
\ba \label{4pde18}
\frac{\partial^2}{\partial \tau_l \partial \tau_m} F_{[f,T_i]}(\tau_i,x)= \frac{\partial^2}{\partial \tau_m \partial \tau_l} F_{[f,T_i]}(\tau_i,x)  \q 
\ea    
at least on the constraint hypersurface. 

Since the partial derivative of $F_{[f,T_i]}(\tau_i,x)$ is again a complete observable, we can apply equation (\ref{4pde10}) to obtain the second partial derivatives of $F_{[f,T_i]}(\tau_i,x)$:
\ba
\frac{\partial^2}{\partial \tau_l \partial \tau_m} F_{[f,T_i]}(\tau_i,x)&=&\frac{\partial}{\partial \tau_l}F_{[g_m,T_i]}(\tau_i,x) \nn \\ &=& 
A^{-1}_{lj}\{C_j,A^{-1}_{mk}\{C_k,f\}\}({\bf T}_x^{-1}(\tau_i)) \nn \\
&=&\left(A^{-1}_{lj}\{C_j,A^{-1}_{mk}\}\{C_k,f\}+A^{-1}_{lj}A^{-1}_{mk}\{C_j,\{C_k,f\}\}\right)({\bf T}_x^{-1}(\tau_i))\q\q
\ea
where here and in the following we sum over repeated indices. Multiplying both sides of this equation with $A_{il}A_{hm}$ gives
\ba \label{4pde20}
A_{il}A_{hm}\frac{\partial^2}{\partial \tau_l \partial \tau_m} F_{[f,T_i]}(\tau_i,x)&=& 
\left(A_{hm}\{C_i,A^{-1}_{mk}\}\{C_k,f\}+\{C_i,\{C_h,f\}\}\right)({\bf T}_x^{-1}(\tau_i)) \nn \\
&=&
\left(-A^{-1}_{mk}\{ C_i,A_{hm}  \}\{C_k,f\}+\{C_i,\{C_h,f\}\}\right)({\bf T}_x^{-1}(\tau_i)) \q\q
\ea
where we used that
\ba
0=\{C_i,\delta_{hm}\}=\{C_i,A_{hm} A^{-1}_{mk}\}=A^{-1}_{mk}\{ C_i,A_{hm}  \}+A_{hm}\{C_i,A^{-1}_{mk}\}  \q .
\ea
The anti-symmetrization of equation (\ref{4pde20}) in the indices $i,h$ is
\ba
A_{[i|\,l\,|}A_{h]m}\frac{\partial^2}{\partial \tau_l \partial \tau_m} F_{[f,T_i]}(\tau_i,x)&=&\tfrac{1}{2}\big(-A^{-1}_{mk}\{ C_i,A_{hm}  \}\{C_k,f\}-
A^{-1}_{mk}\{ C_h,A_{im}  \}\{C_k,f\}   \nn \\
   &&   \q\q
+\{C_i,\{C_h,f\}\}+ \{C_h,\{C_i,f\}\}  \big)({\bf T}_x^{-1}(\tau_i)) \nn \\
&=&
 \tfrac{1}{2}\big(-A^{-1}_{mk} \{C_k,f\}  \left(\,\{ C_i,\{C_h,T_m\}  \}-
\{ C_h,\{C_i,T_m\}  \}\,\right)  \nn \\
   &&   \q\q
+\{C_i,\{C_h,f\}\}- \{C_h,\{C_i,f\}\}  \big)({\bf T}_x^{-1}(\tau_i)) \nn \\
&=&\tfrac{1}{2}\big(-A^{-1}_{mk} \{C_k,f\}\{T_m,\{C_h,C_i\}\}+
\{f,\{C_h,C_i\}\} \big)({\bf T}_x^{-1}(\tau_i))  \nn \\
\ea
where in the second line we used $A_{jm}=\{C_j,T_m\}$ and in the last line we used the Jacobi identity for the Poisson brackets.

Now we apply the first class property of the constraints, i.e. that their algebra closes (on the constraint hypersurface): $\{C_h,C_i\}=f_{hij}C_j$ and use again that $A_{jm}=\{C_j,T_m\}$ 
\ba
A_{[i|\,l\,|}A_{h]m}\frac{\partial^2}{\partial \tau_l \partial \tau_m} F_{[f,T_i]}(\tau_i,x)&=&\big(-A^{-1}_{mk} \{C_k,f\}\left( f_{hij}\{T_m,C_j\}+C_j\{T_m,f_{hij}\}\right) 
 \nn \\ 
&& \q\q + f_{hij}\{f,C_j\}+C_j\{f,f_{hij}\}\,\, \big)({\bf T}_x^{-1}(\tau_i))  \nn \\
&=& \big(A^{-1}_{mk} \{C_k,f\}A_{jm}f_{hij}-A^{-1}_{mk} \{C_k,f\}C_j\{T_m,f_{hij}\} 
\nn \\
&& \q\q + f_{hij}\{f,C_j\}+C_j\{f,f_{hij}\}\,\, \big)({\bf T}_x^{-1}(\tau_i))
\nn \\
&=&\big(C_j \left(-A^{-1}_{mk} \{C_k,f\}\{T_m,f_{hij}\}+\{f,f_{hij}\}\,\right) \,\big)({\bf T}_x^{-1}(\tau_i)) \nn \\
&\simeq& 0 \; .\q\q
\ea
Hence (since from our assumption it follows that $A_{hm}$ is an invertible matrix) the integrability condition for the PDE's (\ref{4pde10}) are satisfied everywhere in $\cm$ for constraint algebras with structure constants and at least on the constraint hypersurface for constraint algebras with structure functions.
\vspace{0.4cm}

{\bf Example 5:}
Here we will consider a kind of deformed $SO(3)$ algebra given on the phase space $\Rl^3_q\times\Rl^3_p$ by
\ba
C_1&=&q_2^{n_2}p_3^{m_3}-q_3^{n_3}p_2^{m_2} \nn \\
C_2&=&q_3^{n_3}p_1^{m_1}-q_1^{n_1}p_3^{m_3} \nn \\
C_3&=&q_1^{n_1}p_2^{m_2}-q_2^{n_2}p_1^{m_1}   \q 
\ea
that is $C_i=\sum_{jk}\eps_{ijk} q_j^{n_j}p_k^{m_j}$. A special case of this example was quantized in \cite{mcp2}. The constraint algebra is given by
\ba
\{C_i,C_j\}=\sum_k \eps_{ijk} \, n_k m_k\, q_k^{n_k-1}p_k^{n_k-1}\, C_k \q ,
\ea
hence it is a first class algebra with structure functions.
The set $\{C_1,C_2,C_3\}$ is not an independent set of constraints: because of the anti-symmetry of $\eps_{ijk}$ we have the relations
\ba
q_1^{n_1}C_1+q_2^{n_2}C_2+q_3^{n_3}C_3=p_1^{m_1}C_1+p_2^{m_2}C_2+p_3^{m_3}C_3=0 \q .
\ea
We will therefore choose as a set of independent constraints $\{C_1,C_2\}$. 

Now we choose as clock variables the functions $T_1=q_1$ and $T_2=q_2$. The PDE's (\ref{4pde10}) for $q_3(\t_1,\t_2)=F_{[q_3,T_1,T_2]}(\t_1,\t_2,\cdot)$ are
\begin{xalignat}{2} \label{deformed29}
&\frac{\partial}{\partial \tau_1}q_3(\t_1,\t_2)=-\frac{m_3\, p_3^{m_3-1} q_1^{n_1}}{m_1\, p_1^{m_1-1} q_3^{n_3}}(\t_1,\t_2) 
& &
\frac{\partial}{\partial \tau_2}q_3(\t_1,\t_2)=-\frac{m_3\, p_3^{k_3-1} q_2^{n_2}}{m_2\, p_2^{m_2-1} q_3^{n_3}}(\t_1,\t_2) 
\end{xalignat}
where we abbreviated $x_i(\t_1,\t_2)=F_{[x_i,T_1,T_2]}(\t_1,\t_2,\cdot)$ with either $x_i=q_i$ or $x_i=p_i$. Since $q_1,q_2$ are the clock variables we have $q_1(\tau_1,\t_2)=\t_1$ and $q_2(\t_1,\t_2)=\t_2$. This leaves us with the unknown functions $p_i(\t_1,\t_2)$ in the PDE's (\ref{deformed29}). Hence we consider also the partial derivatives of these functions:
\begin{xalignat}{2} \label{deformed30}
& \frac{\partial}{\partial \t_1}p_1(\t_1,\t_2)=\frac{n_1\, q_1^{n_1-1} p_3^{m_3}}{m_1\, p_1^{m_1-1} q_3^{n_3}}(\t_1,\t_2) 
&&
\frac{\partial}{\partial \t_2}p_1(\t_1,\t_2)=0
\nn \\
& \frac{\partial}{\partial \t_1}p_2(\t_1,\t_2)=0 
&&
 \frac{\partial}{\partial \t_2}p_2(\t_1,\t_2)=\frac{n_2\, q_2^{n_2-1} p_3^{m_3}}{m_2\, p_2^{m_2-1} q_3^{n_3}}(\t_1,\t_2)
\nn \\
& \frac{\partial}{\partial \t_1}p_3(\t_1,\t_2)=-\frac{n_3\, p_1}{m_1\, q_3}(\t_1,\t_2)
&&
\frac{\partial}{\partial \t_2}p_3(\t_1,\t_2)=-\frac{n_3\, p_2}{m_2\, q_3}(\t_1,\t_2)  \q .
\end{xalignat}
On the constraint surface we have 
\be \label{deformed31}
\frac{p_3^{m_3}}{q_3^{n_3}}\simeq\frac{p_2^{m_2}}{q_2^{n_2}}\simeq\frac{p_1^{m_1}}{q_1^{n_1}}  \q ,
\ee
hence we can replace in (\ref{deformed30}) the term $p_3^{m_3}/q_3^{n_3}$ according to (\ref{deformed31}). Using also $q_1(\tau_1,\t_2)=\t_1$ and $q_2(\t_1,\t_2)=\t_2$ we obtain
\begin{xalignat}{2}
& \frac{\partial}{\partial \t_1}p_1(\t_1,\t_2)\simeq\frac{n_1\, p_1}{m_1\, \t_1}(\t_1,\t_2) 
&&
\frac{\partial}{\partial \t_2}p_1(\t_1,\t_2)=0
\nn \\
& \frac{\partial}{\partial \t_1}p_2(\t_1,\t_2)=0 
&&
 \frac{\partial}{\partial \t_2}p_2(\t_1,\t_2)\simeq\frac{n_2\,  p_2}{m_2\, \t_2}(\t_1,\t_2)
\end{xalignat}
These equations are easily integrable to
\begin{xalignat}{2} \label{deformed33}
&\frac{p_1^{m_1}}{q_1^{n_1}}(\t_1,\t_2)=\frac{p_1^{m_1}(\t_1,\t_2)}{\t_1^{n_1}}\simeq \frac{p_1^{m_1} (\t_{10},\t_{20})}{\t_{10}^{n_1}} &\q&
\frac{p_2^{m_2}}{q_2^{n_2}}(\t_1,\t_2)=\frac{p_2^{m_2}(\t_1,\t_2)}{\t_2^{n_2}}\simeq \frac{p_2^{m_2}(\t_{10},t_{20})}{\t_{20}^{n_2}} \q .
\end{xalignat}
Hence $F_1:=p_2^{m_2}/q_2^{n_2} \simeq p_1^{m_1}/q_1^{n_1}\simeq p_3^{m_3}/q_3^{n_3}$ is conserved and indeed it commutes weakly with the constraints.
Using the relations (\ref{deformed30}) and the solutions (\ref{deformed33}) the equations (\ref{deformed29}) can be written as
\begin{xalignat}{1}
& \frac{\partial}{\partial \t_1}q_3(\t_1,\t_2)=-\frac{m_3\, p_1}{m_1 p_3}(\t_1,\t_2)=-\frac{m_3}{m_1} \frac{p_1(\t_{10},\t_{20})}{p_3(\t_{10},\t_{20})}\frac{\t_1^{n_1/m_1}}{\t_{10}^{n_1/m_1}}\frac{q_3^{n_3/m_3}(\t_{10},\t_{20})}{q_3^{n_3/m_3}(\t_{1},\t_{2})}
\nn \\
& \frac{\partial}{\partial \t_2}q_3(\t_1,\t_2)=-\frac{m_3\, p_2}{m_2 p_3}(\t_1,\t_2)=-\frac{m_3}{m_2} \frac{p_2(\t_{10},\t_{20})}{p_3(\t_{10},\t_{20})}\frac{\t_2^{n_2/m_2}}{\t_{10}^{n_2/m_2}}\frac{q_3^{n_3/m_3}(\t_{10},\t_{20})}{q_3^{n_3/m_3}(\t_{1},\t_{2})}
 \q .
\end{xalignat}
The equations can be integrated to
\ba
\frac{q_3^{(n_3/m_3)+1}(\t_1,\t_2)}{(n_3+m_3)q_3^{n_3/m_3}(\t_{10},\t_{20})}p_2(\t_{10},\t_{20})   &\simeq&  \frac{q_3(\t_{10},\t_{20})p_3(\t_{10},\t_{20})}{n_3+m_3}
 -\frac{\t_1^{(n_1/m_1)+1}}{(n_1+m_1)\t_{10}^{n_1/m_1}}p_1(\t_{10},\t_{20})
\nn \\ &&
-\frac{\t_2^{(n_2/m_2)+1}}{(n_2+m_2)\t_{20}^{n_2/m_2}}p_2(\t_{10},\t_{20})
\ea
From here and (\ref{deformed33}) we obtain that 
\ba 
F_2:=\frac{q_1p_1}{n_1+m_1}+\frac{q_2p_2}{n_2+m_2}+\frac{q_3p_3}{n_3+m_3} \q .
\ea
is a Dirac observable. Hence we found two (independent) Dirac observables $F_1,F_2$ which describe the two-dimensional reduced phase space.
\vspace{0.4cm}

Next we will give a formal solution to the system of PDE's (\ref{4pde10}). To this end consider the (formal) power series of $F_{[f,T_i]}(\tau_i,x)$ in the $\tau_i$'s around the point $\tau_i=T_i(y)$ where $y$ is a point in the gauge orbit $\cg_x$ through $x$:  
\ba \label{4pde24}
F_{[f,T_i]}(\tau_i,x)\!\!&=&\!\!\!\!\!\!\sum_{k_1,\ldots,k_n=0}^\infty \frac{1}{k_1!\cdots k_n!}\frac{\partial^{k_1 \cdots k_n}}{\partial^{k_1}\tau_1 \cdots \partial^{k_n}\tau_n} F_{[f,T_i]}(T_i(y),x)\;(\tau_1-T_1(y))^{k_1}\cdots (\tau_n-T_n(y))^{k_n}  
\nn \\
\ea
We know that all the partial derivatives appearing in (\ref{4pde24}) can be written as complete observables associated to some phase space function $g_{(k_1,\ldots,k_n)}$
\ba
\frac{\partial^{k_1 \cdots k_n}}{\partial^{k_1}\tau_1 \cdots \partial^{k_n}\tau_n} F_{[f,T_i]}=:F_{[g_{k_1,\ldots,k_n},T_i]} \q .
\ea
Furthermore we have by definition of a complete observable 
\ba
F_{[g_{k_1,\ldots,k_n},T_i]}(T_i(y),x)=g_{(k_1,\ldots,k_n)}(y)
\ea
so that we can replace the partial derivatives in (\ref{4pde24}) by $g_{(k_1,\ldots,k_n)}(y)$. Because of equation (\ref{4pde10}) these functions are given by 
\ba \label{4pde27}
g_{(k_1,\ldots,k_n)}= (S_1)^{k_1}\circ\cdots\circ (S_n)^{k_n}(f)
\ea
where $S_j$ is the map
\ba\label{4pde28}
S_j:\cc^\infty(\cm)&\rightarrow& \cc^\infty(\cm) \nn \\
h   &\mapsto& A^{-1}_{jl} \{C_l,h\}  \q .
\ea
(Here we made the assumption that $A^{-1}_{jl}$ has smooth matrix entries.) The order of the operators $S_j$ in (\ref{4pde27}) does not matter (at least on the constraint hypersurface) because of the consistency conditions (\ref{4pde18}). 

Therefore the formal solution to the PDE's (\ref{4pde10}) can be written as
\ba\label{4pde29}
F_{[f,T_i]}(\tau_i,x)\!\!&=&\!\!\!\!\!\!\sum_{k_1,\ldots,k_n=0}^\infty \frac{1}{k_1!\cdots k_n!} \, g_{(k_1,\ldots,k_n)}(y)\;(\tau_1-T_1(y))^{k_1}\cdots (\tau_n-T_n(y))^{k_n}  
\ea
where the functions $g_{(k_1,\ldots,k_n)}(y)$ are defined in (\ref{4pde27},\ref{4pde28}) and $y$ is a point on the gauge orbit $\cg_x$ through $x$.

Assume that the sum (\ref{4pde29}) converges for fixed values of the parameters ${\t_0}_i$ in some neighbourhood of a phase space point $x_0$. Furthermore assume that one can permute differentiation with respect to phase space variables and summation. Then it is straightforward to show directly, that (\ref{4pde29}) poisson commutes (weakly) with the constraints. (Act on (\ref{4pde29}) with the (differential) operator $S_j$ and use that they commute up to terms proportinal to the constraints (which follows from the proof of the consistency conditions (\ref{4pde18}))
\be
S_j\circ S_k (h)(x)=S_k\circ S_j(h)(x)+\lambda_j(x) C_j(x)
\ee
as well as $S_j(T_i)=\delta_{ij}$. The result
\ba
A_{jk}^{-1}\{C_k,F_{[f,T_i]}({\tau_0}_i,x)\}\!\!&\simeq&\!\!\!\!\!\!\sum_{k_1,\ldots,k_n=0}^\infty \frac{1}{k_1!\cdots k_n!} \, S_1^{k_1}\circ\cdots\circ S_j^{k_j+1}\circ S_n^{k_n} (f)(y)\;  \nn \\
&& \q\q\; (\tau_1-T_1(y))^{k_1}\cdots (\tau_n-T_n(y))^{k_n}  \nn \\
&&-\!\!\!\!\!\! \sum_{k_1,\ldots,k_j=1,\ldots, k_n=0}^\infty  \!\! \frac{k_j}{k_1!\cdots k_n!}    S_1^{k_1}\circ\cdots\circ S_n^{k_n} (f)(y)\nn \\
&& \q\q\; (\tau_1-T_1(y))^{k_1}\cdots (\tau_j-T_j(y))^{k_j-1} \cdots(\tau_n-T_n(y))^{k_n} \q\q\q
\ea
vanishes weakly, which can be seen by relabelling the index $k_j$ in the last sum to $k_j'=k_j-1$.)

Hence, if the above mentioned convergence conditions are satisfied, (\ref{4pde29}) provides a local definition of a complete observable (in contrast to theorem \ref{theorem2} where we made global assumptions on the properties of the partial observables with respect to the constraint flow).

\section{Weak Abelianization of the Constraint Algebra} \label{abelian}

In this section we will remark, that one can understand the result (\ref{4pde29}) also from another viewpoint:
The proof for the consistency conditions (\ref{4pde18}) proceeded by showing that 
\ba
A^{-1}_{lj}\{C_j,A^{-1}_{mk}\{C_k,f\}\}-A^{-1}_{mj}\{C_j,A^{-1}_{lk}\{C_k,f\}\}
\ea
vanishes weakly for an arbitrary phase space function $f$. Therefore the flows generated by the vector fields $\tilde{\chi}_m:=A^{-1}_{mj}{\chi_{C_j}}$ commute weakly. Since we have
\ba
\{A^{-1}_{mj}C_j,f\}\simeq A^{-1}_{mj}\{C_j,f\}
\ea
the flows generated by $\tilde{\chi}_m$ and the flows generated by $\chi_{\tilde{C}_m}$ where 
\be \label{Ctilde}
\tilde{C}_m:=A^{-1}_{mj}C_j
\ee
coincide on the constraint hypersurface. Hence also the flows generated by the $\tilde{C}_m$ commute on the constraint hypersurface and indeed one can calculate directly that
$\{\tilde{C}_m,\tilde{C}_j\}$ contains only terms which are at least quadratic in the constraints. We will call this property of the constraint set $\{\tilde{C}_m;m=1,\ldots,n\}$ weakly abelian. (That is, a constraint set is weakly abelian if the associated structure functions vanish weakly.)

Moreover the evolution of the clock variables $T_j$ with respect to these new constraints is linear in the evolution parameters (again restricted to the constraint hypersurface):
\ba
\a_{\b_j\tilde{C}_j}(T_k)(x)\simeq T_k(x)+\delta_{kj}\b_j   \q .
\ea
Now we can equally well use the constraints $\tilde{C}_m$ in the definition of a complete observable:
\ba
\tilde{F}_{[f;T_1,\ldots,T_n]}=\a_{\b_j\tilde{C}_j}(f)_{|\a_{\b_j\tilde{C}_j}(T_k)(x)=\tau_k} \q .
\ea
Then the complete observables $\tilde{F}_{[f;T_1,\ldots,T_n]}$ and $F_{[f;T_1,\ldots,T_n]}$ coincide weakly. The advantage in using the constraints $\tilde{C}_m$ is that now the solution of the equations
\ba
\a_{\b_j\tilde{C}_j}(T_k)(x)=\tau_k
\ea
is very easy, namely given by $\beta_j=\tau_j-T(x)$ for $x$ on the constraint hypersurface. From this we can conclude
\ba\label{4pde35}
\tilde{F}_{[f,T_1,\ldots,T_n]}(\tau_i,x)&=&\a_{\b_j\tilde{C}_j}(f)(x)_{|\beta_j=\tau_j-T(x)} 
\nn  \\
&\simeq& \a_{\b_1\tilde{C}_1}\circ \cdots \circ \a_{\b_n\tilde{C}_n}(f)(x)_{|\beta_j=\tau_j-T(x)} 
\nn \\
&=&\!\!\!\!\!\!\!\!\sum_{k_1,\ldots,k_n=0}^\infty \!\!\frac{1}{k_1!\cdots k_n!} \! \tilde{S}_1^{k_1}\circ\cdots\circ\tilde{S}_n^{k_n}\,(f)(x)\;(\tau_1\!-\!T_1(x))^{k_1}\!\cdots (\tau_n\!-\!T_n(x))^{k_n} \q \q \q
\ea
where 
\ba\label{4pde36}
\tilde{S}_j:\cc^\infty(\cm)&\rightarrow& \cc^\infty(\cm) \nn \\
h   &\mapsto& \{A^{-1}_{jl}C_l,h\} =\{\tilde C_j,h\} \q .
\ea
Indeed formulas (\ref{4pde35}) and (\ref{4pde24}) for the power series of $\tilde{F}_{[f,T_i]}$ and $F_{[f,T_i]}$ respectively coincide on the constraint hypersurface.

More generally, assume that one has found a set of constraints $\{\hat{C}_j,\,j=1,\ldots,n\}$ which has the property, that it evolves the clock variables linearly on the constraint hypersurface, i.e.
\ba \label{weaklyconj}
\{\hat{C}_j,T_k\}=\delta_{jk}+\lambda_{jkm}\hat{C}_m
\ea
where $\lambda_{jkm}$ are smooth phase space functions\footnote{That a function $g$ which vanishes on the constraint hypersurface can always be written as $\lambda_{jkm}\hat{C}_m$ is proven in \cite{henneaux}.}. For such a constraint set one can prove that it is weakly abelian: One either uses that the matrix $\hat{A}_{jk}:=\{\hat C_j,T_k\}$ coincides weakly with the identity matrix and hence also its inverse coincides weakly with the identity matrix. Therefore 
\be
\tilde{\hat{C}}_j:=\hat{A}^{-1}_{jk}\hat C_l=   \hat C_j+\lambda'_{jkl}\hat C_k \hat C_l 
\ee 
for certain phase space functions $\lambda'_{jkl}$. Since we know that $\tilde{\hat{C}}_j$ are weakly abelian, we can conclude that the $\hat C_j$'s are also weakly abelian. An alternative proof proceeds by calculating $\{\hat C_j,\{\hat C_i,T_k\}\}$ firstly directly and then 
by using the Jacobi identity. One then compares the two results and concludes that the structure functions $\hat f_{ijk}$ defined by $\{\hat C_i,\hat C_j\}=\hat f_{ijk} \hat C_k$ vanish weakly.

On the other hand we can also argue that if there exist a set of phase space variables $\{T_1,\ldots,T_n\}$ such that (\ref{weaklyconj}) holds with respect to a constraint set $\{\hat C_1,\ldots,\hat C_n\}$, then the constraints have to be weakly abelian.

The arguments made above show that in the formal power series of a complete observable (\ref{4pde35}) one can replace the constraints $\tilde C_j$ by the constraints $\hat C_j$ as long as the latter have the property (\ref{weaklyconj}) with respect to the clock variables. The resulting complete observable will at least weakly coincide with the original (\ref{4pde35}) (or(\ref{4pde24}) one.

Applying the inverse of the matrix $A_{jk}=\{C_j,T_k\}$ to the constraints $C_j$ provides one way to obtain such a set of constraints. Another way is to construct (strongly) abelian constraints in a way which is explained in \cite{henneaux}. We will repeat this construction here for completeness: To obtain abelian constraints is locally always possible and proceeds in the following way. Assume that the clock variables commute\footnote{In \cite{henneaux} the abelianization is obtained in a slightly more general setting, in which the $T_j$'s are just part of a new set of canonical coordinates, that is the $T_j$'s may also contain conjugated variables: $\{T_j,T_{j'}\}=\pm 1$ or $\{T_j,T_{j'}\}=0$.
}.
Then it is possible to use these clock variables $T_j$ as a part of a new set of canonical coordinates, such that $\Pi_j$ are the momenta conjugated to $T_j$. Call the remaining new canonical coordinates $Y_m, m=1,\ldots 2p-2n$. 

If $\text{det}(A_{jk})=\text{det}(\{C_j,T_k\})\neq 0$ it is in principle possible to solve the constraints locally for the momenta $\Pi_j$, that is the vanishing of the constraints is equivalent to
\ba
\Pi_j=E_j(T_k,Y_m)  \q  .
\ea 
Hence an equivalent set of constraints is given by
\ba
\hat C_j=E_j(T_k,Y_m)-\Pi_j  \q  .
\ea
Obviously we have $\{\hat C_j,T_k\}=\delta_{jk}$ (strongly) and one can also show that the $\hat C_j$ Poisson commute strongly: Since $\hat C_j$ are first class constraints their Poisson brackets vanishes on the constraint hypersurface. But
\ba \label{strongabel}
\{\hat C_i,\hat C_j\}=\{E_i(T_k,Y_m),E_j(T_k,Y_m)\}-\{E_i(T_k,Y_m),\Pi_j\}-\{P_i,E_j(T_k,Y_m)\}
\ea
does not depend on the values of the momenta $\Pi_k$. Hence the Poisson bracket (\ref{strongabel}) has to vanish not only on the constraint hypersurface $\cc$ but on the whole phase space $\cm$.

As we will see in the next section, such a constraint set $\{\hat C_j,\,j=1,\ldots,n\}$ was used before by Kucha\v{r} in the construction of the Bubble-Time Canonical Formalism \cite{bubble}.

\section{Bubble Time Formalism} \label{bubble}

In this section we will explain how the theory of partial observables connects to the Bubble-Time Canonical Formalism introduced by Kucha\v{r} in \cite{bubble}. There the Bubble-Time Formalism was introduced for General Relativity, i.e. a field theory with a totally constrained Hamiltonian. However it is straightforward to apply this formalism to other first class constraint systems.

Given a system on a $2p$-dimensional phase space $\cm$ with $n$ constraints the Bubble-Time Formalism starts from the following assumption: There exists canonical coordinates\\ $(Q_1,\ldots,Q_{(p-n)},P_1,\ldots,P_{(p-n)}; T_1,\ldots,T_n,\Pi_1,\ldots,\Pi_n)$ such that
\begin{itemize}
\item[]
the pairs $(T_i,\Pi_i)$ and $(Q_k,P_k)$ are canonically conjugate;
\item[]
the determinant of $(\{C_j,T_i\})_{j,i=1}^n$ does not vanish on the constraint hypersurface;
\item[]
the constraint equations $C_j=0;\,j=1,\ldots,n$ can be solved for the momenta $\Pi_k;\,k=1,\ldots,n$.
\end{itemize}

Since according to the last assumptions the constraints can be solved for the momenta $\Pi_i=E_i(Q_k,P_k,T_j)$ such that
\be
C_j(Q_k,P_k,T_i,\Pi_h=E_h(Q_k,P_k,T_j))\equiv 0
\ee
we can replace the set $\{C_j,j=1,\ldots,n\}$ by an equivalent set of constraints defined by
\be
\hat C_j=E_j(Q_k,P_k,T_i)-\Pi_j  \q .
\ee

As explained in the last section
 the flow generated by these constraints commutes (strongly) and the clock variables $T_j$ evolve linearly with respect to these flows:
\be
\a_{\b_j \hat C_j}(T_h)=T_h+\b_h  \q  .
\ee 

Consider the flow of the constraint $H:=\b_j \hat C_j$. For either $X_l=Q_l$ or $X_l=P_l$ we have the equations
\ba
\frac{d}{dt} \a_H^t(X_l)&=&\b_j \,\a^t_H(\{\hat C_j,X_l\})= \b_j \, \a^t_H(\{ E_j,X_l\})  \\
\frac{d}{dt} \a_H^t(T_k)&=&\b_j \, \a^t_H(  \{\hat C_j,T_k\})=\b_k   \q . 
\ea
That is we can write
 \ba   \label{bubble6}
\frac{d}{dt} \a_H^t(X_l) =\a^t_H(  \{ E_j,X_l\})  \frac{d}{dt} \a_H^t(T_k)  \q  .
\ea

In \cite{bubble} Kucha\v{r} introduces a new Poissonbracket $[\cdot,\cdot]$, which is defined by using $(Q_k,P_k), k=1,\ldots,p-n$ as a complete set of canonical variables.
One can then replace $\{E_j,X_l\}$ with $[E_j,X_l]$.

Varying the $\b_j$ in $H=\b_j \hat C_j$ one can evolve the phase space functions $X_l$ from the initial data, $Q_k,P_k,T_i$ (fulfilling the constraint equations)
until one reaches $T_h=\tau_h,\,h=1,\ldots,n$. By definition the result will be equal to our complete observables
\ba
F_{[X_l,T_h]}(\tau_h, x=(Q_k,P_k,T_i,\Pi_i=E_i))=\a_{\b_j \hat{C}_j}(X_k)(x)_{|\b_j=T_j(x)-\tau_j} \q  .
\ea   
Hence the Bubble Time Formalism can be reinterpreted using the concepts of complete and partial observables. From equation (\ref{bubble6}) one can again conclude that the complete observables have to satisfy the partial differential equations
\ba \label{bubble8}
\frac{\partial}{\partial \tau_j} F_{[X_l,T_h]}=F_{[\{E_j,X_l\},T_h]}  \q  .
\ea
As alreday mentioned one can replace the brackets $\{\cdot,\cdot\}$ in (\ref{bubble8}) by the new Poisson brackets $[\cdot,\cdot]$ and this is how equation (\ref{bubble8}) appears in \cite{bubble}. (Also there, the integrability conditions for (\ref{bubble8}) are proven.)
However if one replaces there $X_l$ by a function $f$, equation (\ref{bubble8}) is in general only valid (with the new Poisson brackets) if $f$ does not depend on $T_k$ or $\Pi_k$. Otherwise one has to use instead of $[E_l,f]$ the function $\{\hat{C}_l,f\}$. On the other hand since $F_{[f,T_i]}(\tau_i,x)$ gives the value of $f$ at that point in the gauge orbit of $x$ (on the constraint hypersurface) at which the functions $T_i$ give the values $\t_i$ we have
\ba
F_{[f,T_i]}(\tau_i,\cdot)\simeq F_{[f',T_i]}(\tau_i,\cdot)
\ea
where $f'(Q_k,P_k)=f(Q_k,P_k,T_j=\tau_j,\Pi_j=E_j(Q_k,P_k,\tau_h))$. 
Hence it is not necessary to consider functions which depend on $T_i$ or $\Pi_i$.

In \cite{Torre1} Torre realized that $F_{[X_l,T_h]}(\tau_j,\cdot)$ is a Dirac observable for arbitrary values of $\tau_j$. 
He chooses $\tau_j=0$ and calculates 
 a complete set of Dirac observables $F_{[X_l,T_h]}(\tau_j=0,\cdot)$ for the example of cylindrical symmetric waves. We will dicuss this example in more detail in section \ref{fields}.

\section{Gauge Fixing and Dirac Brackets}\label{gauge}

The complete observables $F_{[f,T_j]}(\tau_j,\cdot)$ can also be understood using $T_j(x)-\tau_j\simeq0, j=1,\ldots,n$ as gauge fixings. 
For this the clock variables $T_j$ have to satisfy the following conditions (taken from \cite{henneaux}), describing good gauges:
\begin{itemize}
\item[]
The chosen gauge must be accessible from an arbitrary point on the constraint hypersurface. That is to each point $x$ on the constraint hypersurface there exists a flow of the form $\a_{\b^1_jC_j}\circ\cdots\circ\a_{\b^l_jC_j}$ with $l$ arbitrary, that maps $x$ to a point $y$ satisfying $T_i(y)=\t_i$ for $i=1,\ldots,n$.

\item[]
The conditions $T_i(x)\simeq \t_i$ must fix the gauge completely, that is there is no gauge transformation other than the identity, that preserves $T_i(x)=\t_i$. Locally this means that
\be
\veps_j \{C_j,T_k-\tau_k\}=\veps_j A_{jk} \simeq 0 \q \text{for} \q k=1,\ldots,n
\ee
have the unique solutions $\veps_j= 0,\, j=1,\ldots,n$. Hence $\text{det}(A_{jk})$ has to be non-vanishing on the constraint hypersurface.
\end{itemize}

Obviously we have, that
\be
F_{[f,T_j]}(\tau_j,x)_{|T_j(x)=\tau_j}\simeq f(x)_{|T_j(x)=\tau_j} \q ,
\ee
that is the partial observable $f$ and the associated complete observable $F_{[f,T_j]}(\tau_j,x)$ coincide on the hypersurface defined by the constraints and the gauge conditions. 
Now, given a gauge satisfying the above mentioned conditions, one can find to each gauge restricted phase space function $f_{|T_j=\tau_j}$ a gauge-invariant extension $F$ away from the gauge conditions, which is uniquely defined at least on the constraint hypersurface (see \cite{henneaux, smolin} where this idea is expressed). 

This extension is (weakly) unique, since through each point $x$ of the constraint hypersurface there is given a gauge orbit $\cg_x$ and on each gauge orbit $\cg_x$ there exits exactly one point $y$ with $T_i(y)=\tau_i$, i.e. which satisfies the gauge conditions. 
We will call this point $y={\bf T}_x^{-1}(\tau_j)$. Since the extension has to be gauge-invariant, it has to be constant along each of the gauge-orbits $\cg_x$. The value of $F$ on such a gauge orbit is determined by the gauge restriction f, that is
\ba
F(x)=f({\bf T}_x^{-1}(\tau_j)) \q  .
\ea
Again, we recognize our complete observable $F(x)\simeq F_{[f,T_i]}(\tau_i,x)$. Hence complete observables are simply gauge invariant extensions of gauge restricted functions, as expressed in \cite{henneaux}. This shows also that
\ba
F_{[f,T_j]}(\tau_j,x)\simeq F_{[g,T_j]}(\tau_j,x)
\ea
if the gauge restrictions of $f$ and $g$ coincide because the gauge invariant extension is unique if the above assumptions hold.

As is well known, the symplectic structure induced on the hypersurface $\{C_j=0,\, T_i=0;\, j,i=1,\ldots,n\}$ from the symplectic structure of the phase space $\cm$ is given by the Dirac bracket $\{\cdot,\cdot\}^*$.
Interestingly the Dirac brackets also appear if one calculates the Poisson bracket of two complete observables $F_{[f,T_i]}(\tau_i,\cdot)$ and $F_{[g,T_i]}(\tau_i,\cdot)$. (This has been stated in \cite{henneaux} but an explicit proof is not available in the literature to the best knowledge of the author.)
To explain this in more detail, we will firstly define the Dirac bracket (following \cite{henneaux}).

Consider the matrix $(B_{jk})_{j,k=1}^{2n}$ defined by 
\ba
B_{jk}:=\{\chi_j,\chi_k\}  \q\q\text{with}\q
\chi_j:=
\begin{cases}
 C_j \q &\text{for}\; 1 \leq j \leq n \\
 (T_{j-n}-\t_{j-n}) \q &\text{for}\; n \leq j \leq 2n \q .
\end{cases}
\ea
On the constraint hypersurface the inverse of $B_{jk}$ is given by
\begin{xalignat}{2}
& (B^{-1})_{hl}\simeq \sum_{i,m=1}^n A^{-1}_{ih}A^{-1}_{ml}\{T_i,T_m\}
&&
  (B^{-1})_{h(n+l))}\simeq -A^{-1}_{lh}  \nn \\
& (B^{-1})_{(h+n)l}\simeq A^{-1}_{hl}
&&
  (B^{-1})_{(h+n)(n+l)}\simeq 0  
\end{xalignat}
where $1 \leq h,l \leq n$ and $A_{hl}=\{C_h,T_l\}$.
The Dirac bracket is then given by
\ba \label{6gauge7}
\{f,g\}^* &=&      \{f,g\}- \sum_{j,k=1}^{2n} \{f,\chi_j\}(B^{-1})_{jk}\{\chi_k,g\}  
\nn \\
          &\simeq&   \{f,g\}-  \sum_{h,l=1}^n \{f,\tilde{C}_h\} \{T_h,T_l\}\{\tilde{C}_l,g\}  
+\sum_{h=1}^n \{f,\tilde{C}_h\}\{T_h,g\} 
-\sum_{h=1}^n\{f,T_h\}\{\tilde{C}_h,g\}   \; .\q\q\;
\ea
However there is an alternative way to define the Dirac bracket on the gauge fixed surface (see \cite{henneaux}): For $f$ an arbitrary phase space function define 
\ba \label{6gauge8}
f^*=f-\{f,\chi_j\}(B^{-1})_{jk} \,\chi_k   \q .
\ea
Then we have that
\ba
\{C_i,f^*\}_{|T_l=\t_l}\simeq 0  \q \text{and} \q \{T_i,f^*\}_{|T_l=\t_l}\simeq 0  \q .
\ea
The first of these equations means that $f^*$ Poisson commutes with the constraints to the zeroth order in $(T_j-\t_j)$. Moreover $f^*$ and $F_{[f,T_i]}(\t_i,x)$ coincide weakly on the gauge fixed surface, hence $f^*$ and $F_{[f,T_i]}(\t_i,x)$ coincide up to the first order in $(T_j-\t_j)$. (The functions have the same zeroth order and commute with the constraints at least to the zeroth order. One can also calculate (\ref{6gauge8}) explicitly and compare it to the power series of $F_{[f,T_i]}$ in (\ref{4pde29}).) From this one can conclude that
\be
\{f^*,g\}_{|T_l=\t_l}\simeq\{F_{[f,T_i]}(\t_i,\cdot),g\}_{|T_l=\t_l} 
\ee
for an arbitrary phase space function $g$.

Now it is straightforward to compute that
\ba
\{f^*,g\}_{|T_l=\t_l}\simeq \{f^*,g^*\}_{|T_l=\t_l}\simeq\{f,g\}^*_{|T_l=\t_l} \q .
\ea
Therefore the Poisson and Dirac bracket between two complete observables on the gauge fixed surface is given by
\ba
\{F_{[f,T_i]}(\t_i,\cdot),F_{[g,T_i]}(\t_i,\cdot)\}_{|T_l=\t_l}\simeq \{f,g\}^*_{|T_l=\t_l} \simeq \{F_{[f,T_i]}(\t_i,\cdot),F_{[g,T_i]}(\t_i,\cdot)\}^*_{|T_l=\t_l}    \q .
\ea
Because we know that the Poisson bracket of two gauge invariant functions is again gauge invariant, we can conclude that the Poisson bracket of two complete observables associated to the functions $f$ and $g$ respectively is the gauge invariant extension of the gauge restricted result, that is the complete observable associated to $\{f,g\}^*$. Moreover, as one can directly verify using formula (\ref{6gauge7}), the Dirac bracket of two gauge invariant functions is weakly equal to their Poisson bracket. Hence
\ba
\{F_{[f,T_i]}(\t_i,\cdot),F_{[g,T_i]}(\t_i,\cdot)\}(x)\simeq \{F_{[f,T_i]}(\t_i,\cdot),F_{[g,T_i]}(\t_i,\cdot)\}^* (x)    \simeq F_{[\{f,g\}^*,T_i]}(\t_i,x)  \q .
\ea
That is the map ${\bf F}_{[T_i]}(\t_i)$ defined by
\ba
{\bf F}_{[T_i]}(\t_i): (\cc^{\infty}(\cm)/\ci(\cm),\{\cdot,\cdot\}^*) & \rightarrow & (\cd(\cm)/\ci(\cm),\{\cdot,\cdot\}) \nn \\
f & \mapsto & F_{[f,T_i]}(\t_i,\cdot) 
\ea
where $\cd((\cm))$ is the space of gauge invariant functions on $\cm$ and $\ci(\cm)$ the ideal of smooth functions vanishing on the constraint hypersurface, is a Poisson algebra homomorphism. 
(
That ${\bf F}_{[T_i]}(\t_i)$ is a homomorphism with respect to multiplication and addition was proved in (\ref{pdecomposed}). 
 The space $\cc^{\infty}(\cm)/\ci(\cm)$ has a well defined Dirac bracket since $\{f,C_h\}^*=0$ for arbitrary phase space functions $f$.
)

The appearence of the Dirac bracket in the Poisson bracket of two complete observables is natural since a complete observable $F_{[f,T_i]}(\t_i,\cdot)$ is determined by the values of $f$ on the submanifold $\ct:=\{T_i(x)=\t_i,\, C_i=0 \;\text{for}\;i=1,\ldots,n\}$. That is, the complete observable is determined by the gauge restriction of $f$. Hence also the Poisson bracket of two complete observables associated to $f$ and $g$ respectiviely must be determined by the restriction of $f$ and $g$ to $\ct$. But the induced Poisson bracket on the submanifold $\ct$ is given by the Dirac bracket (see \cite{henneaux}).

Furthermore we want to note that one can define a non-trivial action of gauge transformations on the space of complete observables. This will generalize the idea of `evolving constants of motion' (see \cite{RovPartObs}) to constraint systems with an arbitrary number of constraints. We will denote this action of $\a_{\gamma_jC_j}$ by $\hat\a_{\gamma_jC_j}$:
\ba \label{gaugeaction1}
\hat\a_{\gamma_jC_j}:\cd(\cm) \ni F_{[f;T_i]}(\t_i;\cdot) \mapsto F_{[\a_{\gamma_jC_j}(f);T_i]}(\t_i;\cdot) \in \cd(\cm)  \q .
\ea
The term on the right hand side can also be written as
\ba
F_{[\a_{\gamma_jC_j}(f);T_i]}(\t_i;x)&=&\a_{\b_jC_j}\circ\a_{\gamma_kC_k}(f)(x)_{|\a_{\b_jC_j}(T_i)(x)=\t_i}
\nn \\
&=& \a_{\b'_jC_j}(f)(x)_{|\a_{\b'_jC_j}\circ\a_{-\gamma_kC_k}(T_i)(x)=\t_i}
\nn \\
&=&F_{[f;\a_{(-\gamma_jC_j)} (T_i)]}(\t_i;x)  
\ea
where the second equation holds because if $B_j=\b_j$ solves $\a_{\b_jC_j}(T_i)(x)=\t_i$ then $\a_{\b'_jC_j}=\a_{B_jC_j}\circ\a_{\gamma_kC_k}$ solves
$\a_{\b'_jC_j}\circ\a_{-(\gamma_kC_k)}(T_i)(x)=\t_i$. Therefore the action $\hat{\alpha}_{\b_jC_j}$ of a gauge transformation on a complete observable (that is the gauge invariant extension of a gauge restricted fuction) changes the gauge restriction from $T_i(x)=\t_i$ to $\a_{-\b_jC_j}(T_i)(x)=\t_i$. 

Moreover
\ba \label{gaugeaction17}
F_{[\a_{\gamma_jC_j}(f);T_i]}(\t_i;x)&=&a_{\gamma_jC_j}(f)({\bf T}_x^{-1}(\t_i))
\nn \\
&=& F_{[f;T_i]}(\t'_i;x)  \q\q \text{with} 
\nn \\
\t'_i &=&\a_{\gamma_j C_j}(T_i)({\bf T}_x^{-1}(\t_l) )  \q\q .
\ea
If we choose to work with the constraints $\tilde{C}_j$ (see \ref{Ctilde}) we obtain
\ba
F_{[\a_{\gamma_j\tilde{C}_j}(f);T_i]}(\t_i;x)\simeq F_{[f;T_i]}(\t_i+\gamma_i;x)  \q  .
\ea

Hence we can conclude that for a Poisson bracket between complete observables with differing values of the parameters $\tau_i$ we have
\ba
\{F_{[f;T_i]}(\t_i;\cdot),F_{[g;T_i]}(\t_i';\cdot)\}\simeq F_{[\{f,\a_{(\t_k-\t'_k)\tilde{C}_k}(g)\}^*;T_i]}(\t_i;\cdot)  \q .
\ea

Equation (\ref{gaugeaction17}) shows that the gauge transformation act merely on the parameters $\tau$ in a complete observable. For instance if we rewrite a Dirac observable $d$ as a complete observable as is done in (\ref{4pdeDirac}) the complete observable does not depend on the parameters $\tau_i$. Therefore the action of a gauge transformation on a complete observable associated to a Dirac observable is trivial. Hence it is also not surprising that the action of a gauge transformation respects the Poisson brackets between two complete observables:

\ba
\{\hat\a_{\gamma_jC_j}\big[F_{[f;T_i]}(\t_i;\cdot)\big],\hat\a_{\gamma_jC_j}\big[F_{[g;T_i]}(\t_i;\cdot)\big]\} &=& 
\{F_{[\a_{\gamma_jC_j}(f);T_i]}(\t_i;\cdot),F_{[\a_{\gamma_jC_j}(g);T_i]}(\t_i;\cdot)\} \nn \\
&=&
\{F_{[f;T_i]}(\t'_i;\cdot),F_{[g;T_i]}(\t'_i;\cdot)\} 
\nn \\
&=&
F_{[\{f,g\}^*;T_i]}(\t'_i;\cdot)
\nn \\
&=&
F_{[\a_{\gamma_jC_j}(\{f,g\}^*);T_i]}(\t_i;\cdot) \nn \\
&=&
\hat\a_{\gamma_jC_j}\big[\{F_{[f;T_i]}(\t_i;\cdot),F_{[g;T_i]}(\t_i;\cdot)\}\big]
\ea
where $\tau'_i$ is the same as in the last line in equation (\ref{gaugeaction17}). Here it does not matter for which values of $\tau_i$ one calculates the Dirac bracket $\{f,g\}^*$ since according to formula (\ref{6gauge7}) the Dirac bracket is independent from the choice of the $\tau_i$'s.

\section{Partially Invariant Partial Observables}\label{partinv}

Here we will examine the following question. Assume that we already found phase space functions which are (weakly, i.e. on the constaint hypersurface) invariant under a subalgebra $\Fc_1:=\{C_m,C_{m+1},\ldots,C_n\}$ of the constraint algebra. Is it then possible to take these partially invariant functions as partial observables and to calculate the associated complete observable wit respect to the remaining constraints $\Fc_2=\{C_1,\ldots,C_m\}\,$?

There are two potential obstacles for this procedure. Firstly, the set $\Fc_2$ does not need to be a subalgebra, i.e. it may happen, that Poisson brackets of constraints taken from this set do involve constraints from the set $\Fc_1$. Hence there is no guarantee that the Hamiltonian vector fields associated to the constraints from $\Fc_2$ integrate to an $m$-dimensional hypersurface.

Secondly, if the subalgebra $\Fc_1$ is not an ideal, the flow $\alpha^t_{C_k}(f)$ of a $\Fc_1$-invariant function $f$ generated by a constraint $C_k \in \Fc_2$ needs not to be $\Fc_1$-invariant. Hence it is not clear, whether the complete observable associated to $\Fc_1$-invariant partial observables is still invariant under $\Fc_1$.

There are two ways to investigate this question, the first is to consider the system of PDE's (\ref{4pde10}), the second is to start from the definition (\ref{2defequ1}) of complete observables. We will begin with the examination of the PDE's (\ref{4pde10}).

Assume that we have chosen $n$ clock variables $T_i,i=1,\ldots,n$ in such a way that the first $m$ clock variables $T_j,j=1,\ldots,m$ are invariant under $\Fc_1$ but that the determinant of $A_{kl}=(\{C_k,T_l\})_{k,l=1}^n$ is nowhere vanishing (on the constraint hypersurface). Then we have $A_{ij}=0$ for $i=m+1,\ldots,n$ and $j=1,\ldots,m$. Moreover the determinant of the submatrices $A'_{kl}=(\{C_k,T_l\})_{k,l=1}^m$ and $A''_{kl}=(\{C_k,T_l\})_{k,l=m}^n$ is also nowhere vanishing. Now we choose another $\Fc_1$-invariant partial observable $f$ and consider the PDE's (\ref{4pde10}) 
\ba \label{6part1}
\sum_{j=1}^n A_{kj}({\bf T}_x^{-1}(\tau_i))   \frac{\partial}{\partial \tau_j}F_{[f,T_i]}(\tau_i,x)=\{C_k,f\}({\bf T}_x^{-1}(\tau_i))=0 \q \text{for}\q k=m+1,\ldots,n \q .
\ea  
Since we have $A_{kj}=0$ for $j=1,\ldots,m$ the summation on the left hand side of the equation reduces to the entries of the submatrix $A''_{kj}$. Since the determinant of  $A''_{kj}$ is nowhere vanishing the unique solution to the equations (\ref{6part1}) is 
\ba
\frac{\partial}{\partial \tau_j}F_{[f,T_i]}(\tau_i,x)=0 \q \text{for} \q j=m+1,\ldots,n \q .
\ea 
Hence the complete observable $F_{[f,T_i]}$ does not depend on the last $(n-m)$ of the parameters $\tau_j$. In the remaining PDE's 
\ba \label{6part3}
\sum_{j=1}^n A_{kj}({\bf T}_x^{-1}(\tau_i))   \frac{\partial}{\partial \tau_j}F_{[f,T_i]}(\tau_i,x)=\{C_k,f\}({\bf T}_x^{-1}(\tau_i)) \q \text{for}\q k=1,\ldots,m \q 
\ea      
the summation on the left hand side reduces to the entries of the submatrix $A'_{kj}$ (because we know that $(\partial/\partial \t_j)F_{[f,T_i]}=0$ for $j>m$). Since the determinant of this submatrix is non-vanishing we have
\ba \label{6part4}
\frac{\partial}{\partial \tau_j}F_{[f,T_i]}(\tau_i,x)=\left(\sum_{k=1}^m A'^{-1}_{jk}
\{C_k,f\}\right)({\bf T}_x^{-1}(\tau_i))\q \text{for}\q j=1,\ldots,m 
\ea
where $A'^{-1}_{jk}$ is the inverse of $A'_{kl}$. Now it could a priori happen, that the right hand side of (\ref{6part4}) depends on the parameters $\{\tau_l;\,l=m+1,\ldots,n\}$ through the argument $({\bf T}_x^{-1}(\tau_i))$. However this is excluded through the consistency conditions (\ref{4pde18}). Another way to see this, is to remember that the right hand side of (\ref{6part4}) is again a complete observable $F_{[g_j,T_i]}(\tau_i,x)$ associated to the partial observable $g_j=A'^{-1}_{jk}\{C_k,f\}$ (where here and in the following we sum over repeated indices with summation range $k=1,\ldots,m$). We will now show that $g_j$ is again a (weakly) $\Fc_1$-invariant phase space function. From this it follows that $F_{[g_j,T_i]}(\tau_i,x)$ does not depend on $\{\tau_l;\,l=m+1,\ldots,n\}$.

The proof for the $\Fc_1$-invariance of $g_j$ is similar to the proof for the consistency conditions (\ref{4pde18}). We have to show that
\ba
\{C_h,g_j\}&=&\{C_h, A'^{-1}_{jk}\{C_k,f\}\}
 \nn \\
&=& A'^{-1}_{jk}\{C_h,\{C_k,f\}\}+\{C_h, A'^{-1}_{jk}\}\{C_k,f\}
\ea
vanishes for $h=m+1,\ldots,n$. Multiplying both sides of this equation with $A'_{lj}$ and using the Jacobi identity and the definition $A'_{lj}=\{C_l,T_j\}$ gives
\ba \label{6part6}
A'_{lj}\{C_h,g_j\}&=& A'_{lj}A'^{-1}_{jk}\{C_h,\{C_k,f\}\}+A'_{lj}\{C_h, A'^{-1}_{jk}\}\{C_k,f\} 
\nn \\
&=& \{C_h,\{C_l,f\}\}-A'^{-1}_{jk}\{C_h, A'_{lj}\}\{C_k,f\}
\nn \\
&=& -\{f,\{C_h,C_l\}\}-\{C_l,\{f,C_h\}\}-A'^{-1}_{jk}\{C_h,\{C_l,T_j\}\}\{C_k,f\} \nn \\
&=& -\{f,\{C_h,C_l\}\}-\{C_l,\{f,C_h\}\}+A'^{-1}_{jk}\left(\, \{T_j,\{C_h,C_l\}\}+ \{C_l,\{T_j,C_h\}\}  \, \right)\{C_k,f\} \nn \\
\ea
Since $f$ and $T_j,j=1,\ldots,m$ are $\Fc_1$-invariant functions the second and last term in the last line of (\ref{6part6}) vanishes (at least weakly). The other terms contain the Poisson brackets 
\ba
\{C_h,C_l\}=\sum_{i=1}^m f_{hli}C_i+\sum_{i=m}^n f_{hli}C_i \q .
\ea
Inserting these two sums into the last line of (\ref{6part6}) and using again the $\Fc_1$-invariance of $f$ and $T_j,j=1,\ldots,m$ the second sum vanishes weakly. Hence we are left with 
\ba
A'_{lj}\{C_h,g_j\}&\simeq&-\{f,f_{hli}C_i\} +A'^{-1}_{jk}\{T_j,f_{hli}C_i\}\{C_k,f\}
\nn \\
&\simeq&-f_{hli} \{f,C_i\}+A'^{-1}_{jk}f_{hli}\{T_j,C_i\}\{C_k,f\} 
\nn \\
&\simeq&-f_{hli} \{f,C_i\}-f_{hli}\{C_i,f\}
\nn \\
&\simeq&0
\ea
where in third line we used $\{T_j,C_i\}=-A'_{ij}$. Hence $g_j$ is weakly $\Fc_1$-invariant. Iterating the argument we find, that all the functions $g_{(k_1,\ldots,k_n)}$ defined in (\ref{4pde27}) are weakly $\Fc_1$-invariant.

This also shows that the complete observable $F_{[f,T_i]}(\tau_i,x)$ is $\Fc_1$-invariant (The (non)-invariance of $F_{[f,T_i]}$ was our second potential obstacle from the beginnig of this section): Concerning this we use the formal power series (\ref{4pde29}) of $F_{[f,T_i]}$ in the $\tau_i's$. Firstly we know that every function $g_{(k_1,\ldots,k_n)}$ with some $k_j>0$ for $j=m+1,\ldots,n$ vanishes at least weakly. Hence all the monomials including powers of $\tau_j$ with $j=m+1,\ldots,n$ disappear in the sum (\ref{4pde29}). Secondly all the $g_{(k_1,\ldots,k_n)}$ are weakly $\Fc_1$-invariant. Hence the whole formal power series (\ref{4pde29}) is weakly $\Fc_1$-invariant. 

As we have seen, it is possible to calculate through the systems of PDE's (\ref{4pde10}) the complete observable associated to a $\Fc_1$-invariant function $f$ and $m$  $\Fc_1$-invariant clock variables $T_i;\,i=1,\ldots,m$. (The result is independent from the choice of the remaining  $(n-m)$ clock variables.) How can one understand this starting from the definition
\ba \label{6part9}
F_{[f,T_1,\ldots,T_m]}(\tau_1,\ldots,\tau_m,x)=\a_{\b_j C_j}(f)(x)_{|\a_{\b_j C_j}(T_i)(x)=\tau_i}
\ea
where $\a_{\b_j C_j}$ is the flow associated to $\b_jC_j=\sum_{j=1}^m \b_jC_j\,$? The first obstacle from the beginning of this section was, that the flow generated by the set of constraints $\Fc_2=\{C_1,\ldots,C_m\}$ does not need to integrate to an $m$-dimensional hypersurface in the phase space $\cm$. However, we can introduce the configuration space of the $\Fc_1$-invariant partial observables $\cn'$, that is the $(m+1)$ dimensional manifold coordinatized by the values of the partial observables $(T_1,\ldots,T_m,f)$. As before the flow $\a_{\b_jC_j}(x)$ of a phase space point $x$ induces a flow in $\cn'$ of the point $(T_1(x),\ldots,T_m(x),f(x))$ by  
\ba
\a_{\b_j C_j}(T_1(x),\ldots,T_m(x),f(x))=(\a_{\b_j C_j}(T_1)(x),\ldots,\a_{\b_j C_j}(T_m)(x),\a_{\b_j C_j}(f)(x) )  \q .
\ea 
We will now show that this flow in $\cn'$ is integrable to an hypersurface. For this purpose we have to check that the infinitesimal flows are in involution, i.e that for $C_i,C_j \in \Fc_2$ \\
\ba
(\a_{\veps_1C_i})^{-1}\circ(\a_{\veps_2C_j})^{-1}\circ\a_{\veps_1C_i}\circ\a_{\veps_2C_j}\!\!\!\!\!\!\!\!\!&&(T_1(x),\ldots,T_m(x),f(x))
\nn \\
&&\simeq\a_{\veps_1\veps_2 \gamma_k C_k}(T_1(x),\ldots,T_m(x),f(x))+\co(\veps^3)
\ea
where we abbreviated $\gamma_k C_k=\sum_{k=1}^m \gamma_k C_k$. (The inverse of a flow is $(\a_{\beta_k C_k})^{-1}=\a_{-\beta_k C_k}$.) This follows from the fact, that for a $\Fc_1$-invariant phase space function $T$ 
\ba 
(\a_{\veps_2C_j})^{-1}\circ(\a_{\veps_1C_i})^{-1}\circ(\a_{\veps_1C_i}\circ\a_{\veps_2C_j})(T) &=&
\veps_1\veps_2 \left(\{C_i,\{C_j,T\}\}- \{C_j,\{C_i,T\}\}\right) + \co(\veps^3)
\nn \\
&=&\veps_1\veps_2 \{T,\{C_j,C_i\}\}+ \co(\veps^3)
\nn \\
&\simeq&  \veps_1\veps_2 \big( \sum_{k=1}^m f_{ijk} \{C_k,T\}+\!\!\!\!\sum_{k=m+1}^n \!\!\!f_{ijk} \{C_k,T\}\big)\!+\! \co(\veps^3)
\nn \\
&\simeq& \a_{\veps_1\veps_2 \gamma_k C_k}(T)+ \co(\veps^3)
\ea
where $ \gamma_k C_k=\sum_{k=1}^m f_{ijk}C_k$. The second sum in the third line vanishes (weakly) because of the $\Fc_2$-invariance of $T$. 

Hence the induced flow in $\cn'$ generated by the set $\Fc_2$ integrates to an $m$ dimensional gauge orbit in $\cn'$. The set $\{T_i;\,i=1,\ldots,m\}$ parametrizes this gauge orbit. The complete observable defined in (\ref{6part9}) gives the value of $f$ (or the $(m+1)$-th coordinate in $\cn'$) at the point of this gauge orbit with the parameter values $(\tau_1,\ldots,\tau_m)$. This provides the reason for the $\Fc_2$-invariance of the complete observable (\ref{6part9}).

The reason for the $\Fc_1$-invariance of the complete observable (\ref{6part9}) is that the induced flow in $\cn'$ generated by a constraint $C_l \in \Fc_1$ leaves the orbits of the set $\Fc_2$ invariant. To see this one has to show that
 a $\Fc_1$-invariant function $T$ satisfies for suitable $\gamma_k$
\ba
\a_{\veps C_l}\circ\a_{\veps_k C_k}(T) \simeq \a_{(\veps_k+\veps \gamma_k) C_k}(T)+\co(\veps^2,\veps \veps_k,\veps_k \veps_{k'})    
\ea
where the sums in $k$ extend from $k=1$ to $k=m$. Here one uses again the Jacobi identity and the $\Fc_1$-invariance of $T$.

Therefore, if we apply $\a_{\veps C_l}$ to the complete observable (\ref{6part9})
\ba
\a_{\veps C_l}\big(F_{[f,T_1,\ldots,T_m]}(\tau_1,\ldots,\tau_m,\cdot)\big)(x)=\a_{\veps C_l}\circ   \a_{\b_j C_j}(f)(x)_{|\a_{\veps C_l}\circ  \a_{\b_j C_j}(T_i)(x)=\tau_i}
\ea
it still prescribes to look for the point of the gauge orbit (in $\cn'$) parametrized by the parameter values $(\tau_1,\ldots,\tau_m)$ and to evaluate $f$ at this point. We conclude that the complete observable (\ref{6part9}) is also $\Fc_1$-invariant.

The partially invariant observables provide an example for the fact, that it is possible to define a complete observable, also if the set of clock variables does not provide a perfect parametrization of the gauge orbit, as long as the partial observable $f$ does not need such a perfect parametrization. In our case the clock variables $\{T_1,\ldots,T_m\}$ are constant along the orbits of the subgroup $\Fc_1$, hence they do not provide a parametrization of these suborbits. But also $f$ is constant along these suborbits.

\section{Field Theories} \label{fields}

Here we will make some comments on partial and complete observables for constrained field theories.

The phase space $\cm$ is some Banach space of fields given on a spatial manifold $\Sigma$ and subject to some boundary conditions. The symplectic structure is defined via canonical coordinates $(\phi_a(\sigma),\pi_a(\sigma);\sigma \in \Sigma)$ where $a$ is from some finite index set $\ca$. The non-vanishing Poisson brackets are given by 
\ba
\{\phi_a(\sigma),\pi_b(\sigma')\}=\delta_{ab}\,\delta(\sigma,\sigma')
\ea
where $\delta(\sigma,\sigma')$ is the delta function on $\Sigma$.

Constraint field theories have an infinte set of constraints $C_\mu(\sigma)$, labelled by an index $\mu$ from some finite index set $\ci$ and by the points $\sigma$ of $\Sigma$. Hence we need infintely many clock variables $T_\mu(\sigma)$, that is $|\ci|$ fields built from the canonical fields $(\phi_a(\sigma),\pi_a(\sigma))$. These fields can also be seen as functionals labelled by $\mu,\sigma$, which map a phase space configuration $(\phi_a,\pi_a)\in \cm$ to the value $T_\mu(\sigma)$.

The equivalent to the partial observable $f$ may be the value of some field $f$ at a certain point $\sigma_0$ or more generally some functional $f: \cm \ni (\phi_a,\pi_a)\mapsto f[\phi_a,\pi_a]\in \Rl$.

Then the interpretation of the complete observable $F_{[f;T_\mu(\sigma)]}(\tau_\mu(\sigma);(\phi_a,\pi_a))$ is the following:
Consider a phase space configuration $(\phi_a,\pi_a)$ and apply arbitrary gauge transformations to it. If a gauge transformation is chosen in such a way that the fields $T_\mu(\sigma)$ coincide with the fields $\tau_\mu(\sigma)$ the functional $f$ assumes the value $F_{[f;T_\mu]}(\tau_\mu(\sigma);(\phi_a,\pi_a))$.

This definition of complete observables for field theories differs drastically from the one used in \cite{RovPartObs}. Whereas we remain strictly in a canonical picture (i.e. we are working with fields given on the {\it spatial} manifold $\Sigma$), the definition in \cite{RovPartObs} works with {\it space-time} entities, i.e. fields on the space-time manifold.  

In background independent field theories on a $d$-dimensional spatial manifold $\Sigma$ the constraint set includes $d\times\infty^d$ diffeormorphism constraints $C^D_i(\sigma),i=1,\ldots,d$ and $1\times\infty^d$ Hamiltonian constraint $C^H$. We will assume in the following that there are no other constraints (or that one already managed to reduce the theory with respect to the other constraints). The Hamiltonian in background independent theories is just the sum of the constraints smeared with arbitrary parameters, called lapse (for the Hamiltonian constraint) and shift functions (for the diffeomorphism constraint):  
\ba
H(N,N_i)=\int_\Sigma \left( N(\sigma)C^H(\sigma)+\sum_{i=1}^d N_i(\sigma)C_i^D(\sigma)\right) d^d\sigma  \q .
\ea

The following example shows again that the definition of complete observables used here is a generalization of Kucha\v{r}'s Bubble-Time Formalism \cite{bubble}:

\vspace{0.4cm}
{\bf Example 7:} 
Here we will discuss a reduced gravitational model, namely cylindrically symmetric gravitational waves, also called Einstein-Rosen waves. This model was treated by Kucha\v{r} in \cite{bubble} as an example for the Bubble Time Formalism. In \cite{Torre1} Torre calculated a complete set of Dirac observables for this model using essentially the concept of a complete observable. Both works used the results obtained by Kucha\v{r} in \cite{kuchar}. So this will be just a short repitition and reinterpretation of these works in the light of partial and complete observables. This example appears here for completeness, for details we refer the reader to \cite{kuchar,bubble,Torre1}. 

Consider a spacetime manifold $\cs$ diffeomorphic to $\Rl\times\Rl^2\times S^1$ with coordinates $(t,r,\phi,z)$, where $t,z\in \Rl;\, r\geq 0$,and $\phi \in [0,2\pi)$. On this manifold allow all metrics of the form
\ba\label{cyl1}
d s^2=e^{\gamma-\psi}\left( \,(-N^2+N_1^2)  \,dt^2+2N_1 \,dt\,dr+dr^2 \right)+R^2 e^{-\psi}\, d\phi^2+e^\psi \,dz^2
\ea
where the lapse $N$, the shift $N_1$ and the metric components $R\geq 0,\gamma,\psi$ are all functions of $r$ and $t$ coordinates only, i.e. constant in $\phi$ and $z$. After specifying boundary conditions which can be found in \cite{kuchar,Torre1} one can plug the metric (\ref{cyl1}) into the Einstein-Hilbert action and perform a canonical analysis. The resulting canonical theory has a phase space which is described by the canonical fields (that is functions of $r$) $(\gamma,R,\psi;\,\Pi_\gamma\Pi_R,\Pi_\psi)$ and is totally constrained where the constraints are given by
\ba \label{cyl2}
C^H &=& \Pi_\gamma\Pi_R+\tfrac{1}{2}R^{-1}\Pi^2_\psi+2R''-\gamma'R'+\tfrac{1}{2}R\psi'^2
\\ \label{cyl3}
C^D &=& -2\Pi'_\gamma+\gamma'\Pi_\gamma+R'\Pi_R+\psi'\Pi_\psi  \q  .
\ea
Here $f'$ denotes the derivative of $f$ with respect to the radial coordinate $r$ (with the exception of $r',r''$ which will denote just different values of the radial coordinate). As shown in \cite{kuchar} there is a canonical transformation which simplifies very much the discussion of this model. It changes the canonical pairs $(\gamma,\Pi_\gamma)$ and $(R,\Pi_R)$ to $(T,P_T)$ and $(R,P_R)$ where
\ba \label{cyl4}
T(r) &=& -\int_\infty^r \Pi_\gamma({r'}) d{r'}
 \nn \\
P_T(r) &=& -\gamma'+\frac{\partial }{\partial r} \ln (R'^2-\Pi_\gamma^2) 
 \nn \\
P_R(r) &=& \Pi_R+\frac{\partial }{\partial r} \ln\! \left( \frac{R'-\Pi_\gamma}{R'+\Pi_\gamma}\right)  \q  .
\ea
What we have in mind is to use the fields $T_0(r):=T(r)$ and $T_1(r):=R(r)$ as clock variables and to compute the complete observables associated to $\psi$ and $\Pi_\psi$. For this it is actually not neccessary to introduce the new canonical coordinates, however it simplifies the calculations considerably.

The constraints (\ref{cyl2},\ref{cyl3}) written in the new canonical coordinates are 
\ba \label{cyl5}
C^H &=& R'P_T+T'P_R+\tfrac{1}{2}(R^{-1}\Pi^2_\psi +R\psi'^2) 
 \\ \label{cyl6}
C^D &=& T'P_T +R'P_R+\psi'\Pi_\psi  \q .
\ea
From (\ref{cyl6}) it follows that $R,T$ and $\psi$ tansform as scalars under a transformation of the coordinate $r$. The constraints (\ref{cyl5},\ref{cyl6}) are the same as for the parametrized canonical formalism for a massless, cylindrically symetric scalar field on a Minkowski background. In this parametrized formalism the functions $(T(r),R(r))$ describe a cylindrically symmetric embedding $\Ft$ of a $3$-dimensional surface $\Sigma$ into $4$-dimensional Minkowski spacetime with cylindrical coordinates $(T,R,\Phi,Z)$:
\ba
\Ft: \Sigma \ni (r,\phi,z)\mapsto (T(r),R(r),\Phi=\phi,Z=z)\in \Rl^4  \q .
\ea

Let us calculate the constraints $\tilde{C}_\mu,\,\mu=0,1$, with respect to which the clock variables evolve linearly, and which is defined by
\ba
\tilde{C}_\mu(r)=\int A^{-1}_{\mu\nu}(r,r')C_\nu(r') \, dr'
\ea
where $C_0:=C^H$ and $C_1=C^D$. Here and in the following we will sum over repeated indices. The matrix $A^{-1}_{\mu\nu}(r,r')$ is the inverse of $A_{\mu\nu}(r,r'):=\{C_\mu(r),T_\nu(r')\}$, that is the matrix satisfies
\ba
\int A^{-1}_{\mu\nu}(r,r') A_{\nu\rho}(r',r'')\,dr'=\delta_{\mu\rho}\,\delta(r,r'') \q .
\ea
One obtains
\begin{xalignat}{2}
& (A^{-1})_{00}(r,r')=\frac{-R'}{R'^2-T'^2}\delta(r,r') & &
  (A^{-1})_{01}(r,r')=\frac{T'}{R'^2-T'^2}\delta(r,r') 
\nn \\
& (A^{-1})_{10}(r,r')=\frac{T'}{R'^2-T'^2}\delta(r,r') &&
  (A^{-1})_{11}(r,r')=\frac{-R'}{R'^2-T'^2}\delta(r,r')
\end{xalignat}
for the matrix entries of $A^{-1}$. This gives for the weakly abelianized constraints $\tilde{C}_\mu$
\ba
\tilde{C}_0&=&-P_T-(R'^2-T'^2)^{-1}\left[ \tfrac{1}{2}R'(R^{-1}\Pi^2_\psi+R\psi'^2)-T''\psi'\Pi_\psi\right]
\nn \\
\tilde{C}_1&=&-P_R-(R'^2-T'^2)^{-1}\left[ -\tfrac{1}{2}T'(R^{-1}\Pi^2_\psi+R\psi'^2)+R'\psi'\Pi_\psi\right] \q .
\ea
As explained in \cite{Torre1} one can obtain the original constraints ${C}_\mu$ from the new constraints $\tilde{C}_\mu$ as the normal and tangential projections of $(\tilde{C}_1,\tilde C_2,0,0)$ to the hypersurface $\Sigma$, embedded in the Minkowski space $\Rl^4$.

Now the PDE's (\ref{4pde10}) change to functional differential equations:
\ba
\frac{\delta}{\delta \tau_\rho(r')} F_{[f;T_\mu(r)]}(\t_\mu(r);x)=F_{[\{\tilde{C}_\rho(r'),f\};T_\mu(r)]}(\t_\mu(r);x)
\ea
where $x$ denotes a phase space point, that is the fields $(T,R,\psi;P_T,P_R,\Pi_\psi)$ on $\Sigma$. As the partial observable $f$ we will choose the fields $\psi(r'')$ and $\Pi_\psi(r'')$ evaluated at the point $r''\in \Sigma$.

In the following we will abbreviate $F_{[f;T_\mu(r)]}(\t_\mu(r);x)$ with $f[\t_0,\t_1]$. The functional differential equations for $\psi(r'')[\t_0,\t_1]$ and $\Pi_\psi(r'')[\t_0,\t_1]$ are given by
\ba \label{cyl13}
\frac{\delta}{\delta \tau_0(r')} \psi(r'')[\t_0,\t_1] &=& (R'^2-T'^2)^{-1}\left(R'R^{-1} \Pi_\psi-T'\psi'\right)(r')[\t_0,\t_1]    \delta(r',r'')
\nn \\
\frac{\delta}{\delta \tau_1(r')} \psi(r'')[\t_0,\t_1] &=& (R'^2-T'^2)^{-1}\left(-T'R^{-1} \Pi_\psi+R'\psi'\right)(r')[\t_0,\t_1]   \delta(r',r'')
\nn \\
\frac{\delta}{\delta \tau_0(r')} \Pi_\psi(r'')[\t_0,\t_1] &=&- (R'^2-T'^2)^{-1}\left(\psi'R'R-T'\Pi_\psi \right)(r')[\t_0,\t_1]  \partial_{r'}\delta(r',r'')
\nn \\
\frac{\delta}{\delta \tau_1(r')} \Pi_\psi(r'')[\t_0,\t_1] &=&- (R'^2-T'^2)^{-1}\left(\psi'T'R+R'\Pi_\psi\right)(r')[\t_0,\t_1]  \partial_{r'}\delta(r',r'')  \q .
\ea
In order to solve these differential equations we will follow \cite{Torre1} and consider the equations for a special choice of the parameter fields $\t_0,\t_1$, namely we will set $\tau_0(r)\equiv t$ and $\tau_1(r)\equiv r$. We then vary $\t_0$ from $\tau_0(r)\equiv t$ to $\tau_0(r)\equiv t +\veps$. This gives the partial derivative with respect to $t$ of $\Psi(t,r''):=\psi(r'')[\tau_0\equiv t,\t_1\equiv r]$ and $\Pi_\Psi(t,r''):=\Pi_\psi(r'')[\tau_0\equiv t,\t_1\equiv r]$:
\ba
\frac{\partial}{\partial t} \Psi(t,r'') &=& \int \frac{\delta}{\delta \tau_0(r')} \psi (r'')[\tau_0\equiv t,\t_1\equiv r]\,dr'\; =\;(r'')^{-1}\Pi_\Psi (t,r'')
\nn \\
\frac{\partial}{\partial t} \Pi_\Psi(t,r'') &=& \int \frac{\delta}{\delta \tau_0(r')} \Pi_\psi (r'')[\tau_0\equiv t,\t_1\equiv r]dr'\; =\;\left(r'' \Psi''+\Psi'\right)(t,r'')  \q .
\ea 
Hence $\Psi(r,t)$ satisfies the cylindricall symmetric wave equation
\ba \label{cyl15}
\frac{\partial^2}{\partial t^2}\Psi(t,r)=\left(\Psi''+r^{-1}\Psi'\right)(t,r) \q .
\ea
The general solution to this wave equation is
\ba \label{cyl16}
\Psi(t,r)=\int_0^\infty \left(a_\omega u_\omega(t,r)+\overline{a_\omega u_\omega}(t,r)\right) d\omega
\ea
where $\overline{v}$ is the complex conjugate of $v$, the $a_\omega$ are complex numbers and the functions $u_\omega$ are given by
\be
u_\omega (t,r)=\sqrt{\tfrac{1}{2}}\,J_0(\omega r)\exp(-i\omega t)
\ee
with $J_0$ the Bessel function of the first kind to the $0$th order.

To obtain the solutions for general parameter fields $\t_0(r),\t_1(r)$ we use that $T_0(r)=T(r)$ and $T_1(r)=R(r)$ describe the embedding of the hypersurface $\Sigma$ into the Minkowski space $\Rl^4$. Changing $\t_0,\t_1$ changes the embedding of the hypersurface into the Minkowski space. But we know the solution for the embeddings $\Sigma \ni (r,\phi,z)\mapsto (T=t,R=r,\Phi=\phi,Z=z)\in \Rl^4$. Therefore we can obtain the general solution of $\psi(r')[\t_0,\t_1]$ by pulling back the solution $\Psi(t,r)$ (equation \ref{cyl16}) from $\Rl^4$ to $\Sigma$:
\ba \label{cyl18}
\psi(r')[\t_0,\t_1]=\Psi(\tau_0(r'),\tau_1(r')) \q  .
\ea
For the general solution $\Pi_\psi(r')[\t_0,\t_1]$ we consider the first two equations in (\ref{cyl13}) and use that $T(r')[\tau_0,\tau_1]=\tau_0(r)$ as well as $R(r')[\tau_0,\tau_1]=\tau_1(r)$. Furthermore we utilize equation (\ref{cyl18}) :
\ba\label{cyl19}
\frac{\delta}{\delta \tau_0(r')} \psi(r'')[\t_0,\t_1] &=& \partial_{\t_0}\Psi(\t_0(r'),\t_1(r'))\delta(r',r'')
\nn \\
&=&(\t_1'^2-\t_0'^2)^{-1} \bigg(\t_1'\t_1^{-1}\Pi_\psi(r')[\t_0,\t_1]-
\nn \\
&&
\t_0'(\partial_{\t_0}\Psi(\t_0(r'),\t_1(r'))\t_0'+\partial_{\t_1}\Psi(\t_0(r'),\t_1(r'))\t_1')\bigg)(r')\delta(r',r'') 
\nn \\
\frac{\delta}{\delta \tau_1(r')} \psi(r'')[\t_0,\t_1] &=& \partial_{\t_1}\Psi(\t_0(r'),\t_1(r'))\delta(r',r'')
\nn \\
&=&(\t_1'^2-\t_0'^2)^{-1} \bigg(-\t_0'\t_1^{-1}\Pi_\psi(r')[\t_0,\t_1]+
\nn \\
&&\t_1'(\partial_{\t_0}\Psi(\t_0(r'),\t_1(r'))\t_0'+\partial_{\t_1}\Psi(\t_0(r'),\t_1(r'))\t_1')\bigg)(r')\delta(r',r'') \q . \nn \\
\ea
These equations can be solved for the momenta $\Pi_\psi(r')[\t_0,\t_1]$:
\ba \label{cyl20}
\Pi_\psi(r')[\t_0,\t_1]=\t_1(r')\bigg(\tau_1'(r') \, \partial_{\t_0}\Psi(\t_0(r'),\t_1(r'))+\t_0'(r') \, \partial_{\t_1}\Psi(\t_0(r'),\t_1(r'))\bigg) \q .
\ea

It remains to incorporate the initial conditions
\ba \label{cyl21}
F_{[\psi(r');T_\mu(r)]}(\t_\mu(r),;x)_{|\{\t_0(r)\equiv T(r),\t_1(r)\equiv R(r)\}}\!\! &=&\!\!\!\! \psi(r')[\t_0(r)\equiv T(r),\t_1(r)\equiv R(r)] \stackrel{!}{=}\psi(r') \nn \\
F_{[\Pi_\psi(r');T_\mu(r)]}(\t_\mu(r),;x)_{|\{\t_0(r)\equiv T(r),\t_1(r)\equiv R(r)\}}\!\! &=&\!\!\!\! \Pi_\psi(r')[\t_0(r)\equiv T(r),\t_1(r)\equiv R(r)]\stackrel{!}{ =} \Pi_\psi(r') \q\q\q \nn \\
\ea
 in order to specify the numbers $a_\omega$ in (\ref{cyl16}) as functions of the phase space point\\ $x=(T,R,\psi;P_T,P_R,\Pi_\psi)$ (where the momenta $P_T,P_R$ are fixed by the constraints (\ref{cyl5},\ref{cyl6})). Since the complete observables do depend on the phase space point $x$ only through the functions $a_\omega(x)$, these have to be Dirac observables (as $\psi(r')[\t_0,\t_1]$ and $\Pi_\psi(r')[\t_0,t_1]$ are Dirac observables for arbitrary parameter fields $\t_0(r)$,$\t_1(r)$ and arbitrary points $r'$). 

Using the solutions (\ref{cyl16},\ref{cyl18},\ref{cyl20}) the initial conditions (\ref{cyl21}) can be written as
\ba
\psi(r')\!\!& \stackrel{!}{=}&\!\!
\sqrt{\tfrac{1}{2}}\int_0^\infty\!\!  \left(a_\omega(x) J_0(\omega R(r'))e^{-i\omega T(r')}+\text{c.c.}\right)d\omega
\nn \\
\Pi_\psi(r') \!\!&\stackrel{!}{=}&\!\!
\sqrt{\tfrac{1}{2}}\int_0^\infty \!\!\left(a_\omega(x)R(r')e^{-i\omega T(r')} \left(-i\omega R'(r')J_0(\omega R(r'))-\omega T'(r')J_1(\omega R(r'))\right)+\text{c.c.} \right) d\omega
\nn \\
\ea
where $J_1(\cdot)=-J_0'(\cdot)$ is the Bessel function of the first kind to the first order. 

For the special phase space points $x^*=(T(r)=t,R(r)=r,\psi(r),\Pi_\psi(r))$ (on the constraint hypersurface) these equations can be solved to
\ba \label{cyl23}
a_\omega(x^*)=\frac{1}{\sqrt{2}}\int_0^\infty \!\!\!\!\! e^{i\omega t}\left(\psi(r)\omega r J_0(\omega r)+ i J_0(\omega r) \Pi_\psi(r)\right) dr \q .
\ea
Here one uses that
\ba
\int_0^\infty  \int_0^\infty b(\omega) J_0(\omega r)J_0(\omega' r) \omega d\omega\, rdr =b(\omega')  \q .
\ea

For general phase space points $x$ we will follow the argumentation in \cite{Torre1} and refer to the fact that the model can be seen as a parametrized (cylindrically symmetric) scalar field on Minkowski space. For such parametrized scalar fields the Klein-Gordon inner product 
\ba
(\psi_1,\psi_2)=i\int_\Sigma q^{1/2} \left(\overline{\psi_1}\, n^\mu \partial_\mu \psi_2 -\psi_2 \, n^\mu \partial_\mu \overline{\psi_1}\right) d\Sigma
\ea 
is independent from the choice of the embedding of $\Sigma$ into the Minkowski space. Here $q$ is the determinant of the induced metric (coming form the Minkowski space) on $\Sigma$ and $n^\mu$ is the normal to this embedded hypersurface. The independence on the embedding can be proved by using Stokes theorem and the equations (\ref{cyl15}) and (\ref{cyl18}).

Now (\ref{cyl23}) can be written as the Klein-Gorden inner product between the mode function $u_\omega(t,r)$ and the field $\psi(r)$. Generalizing this equation to an arbitrary embedding specified by $(T(r),R(r))$ gives
\ba
a_\omega(x)=\frac{1}{\sqrt{2}}\int_0^\infty \!\!\!\!\! e^{i\omega T(r)} \big(\psi(r)\omega R(r)\,[ i T'(r) J_1(\omega R(r))+R'(r)J_0(\omega R(r))]+iJ_0(\omega R(r))\Pi_\psi(r) \big)dr \q  
\nn \\
\ea
as was already computed in \cite{Torre1}. There it is also shown explicitly that the $a_\omega(x)$ are Dirac observables (i.e. that they commute with the constraints).

\vspace{0.4cm}

Finally we will sketch how one could apply the ideas developed in section \ref{partinv} to background independent field theories. Here it is natural to try to work with partial observables which are invariant under spatial diffeomorphisms. This would leave us with the gauge transformations generated by the Hamiltonian constraints. Since these are still $1\times \infty^d$ constraints we also need $1\times \infty^d$ spatial-diffeomorphism invariant clock variables. Hence we need a spatial-diffeomorphism invariant partial observable for each point $\sigma$ in the spatial hypersurface. One could construct those by using for instance a material reference system, see \cite{deWitt}. That is, one would construct spatial-diffeomorphism invariant phase space functions as complete observables associated to $(d+1)$ scalar fields $\psi_1,\ldots \psi_d$ with respect to the spatial-diffeomorphism constraints. 

More concretely one chooses $d$ scalar fields of clock variables $T_i(\sigma)=\psi_i(\sigma)$ and as the partial observable $f=\psi_{d+1}(\sigma^*)$, another scalar field evaluated at a point $\sigma^* \in \Sigma$. One then computes the complete observable $F_{[f,T_i(\sigma)]}(\t_i(\sigma),(\phi_a,\pi_a))$ with respect to the diffeomorphism constraints $C^D(\sigma)$. Consider the definition of such a complete observable:
\ba
F_{[f,T_i(\sigma)]}(\t_i(\sigma),(\phi_a,\pi_a))=\a_{C^D(\beta_j)}\left(\psi_{d+1}(\sigma^*)\right)[\phi_a,\pi_a]_{| \a_{C^D(\beta_j)}\left(\psi_{i}(\sigma)\right)[\phi_a,\pi_a]=\t_i(\sigma)}
\ea
where $\a_{C^D(\beta_j)}$ is the flow generated by the constraint $C^D(\beta_j)=\int  \beta_j(\sigma)C^D_j(\sigma)d^d\sigma$. Now it is known, that the transformation generated by $C^D(\beta_j)$ is simply given by the action of an appropriate spatial diffeomorphism $\varphi_{(\beta_j)}:\Sigma \rightarrow \Sigma$ on the fields. Since the $\psi_k,k=1,\ldots,d+1$ are scalar fields, the diffeomorphism $\varphi_{\b_j}$ acts as a pull back, that is we have
\ba \label{9.31}
F_{[\psi_{d+1}(\sigma^*);T_i(\sigma)]}(\t_i(\sigma),(\phi_a,\pi_a))=\psi_{d+1}(\varphi_{(\b_j)}(\sigma^*))[\phi_a,\pi_a]_{|\psi_i(\varphi_{(\b_j)}(\sigma))[\phi_a,\pi_a]=\t_i(\sigma)}   \q .
\ea 
The meaning of (\ref{9.31}) is the following: Evaluate the scalar field $\psi_{d+1}$ at the point $\sigma^{**}$ which is specified by the condition $\psi_i(\sigma^{**})=\t_i(\sigma^*)=:v_i$.
Varying of the parameters $v_1,\ldots,v_d$ gives the right number of diffeomorphism invariant observables. So for the spatial diffeomorphism group we recover the complete observables used in \cite{RovPartObs}.

To construct a Dirac observable with respect to all the constraints $C^H,C^D$ we have to find another spatial diffeomorphism invariant observable $g$ and to compute the complete observable associated to $g$ and to the spatial diffeomorphism invariant clock variables $\psi_{d+1}(v_1,\ldots,v_d)$ with respect to the Hamiltonian constraints $C^H(\sigma);\, \sigma \in \Sigma$.
Further investigations into this proposal will be undertaken in \cite{fields}.

\section{Discussion}\label{discuss}

Our aim was, to give a construction principle for Dirac observables via the concepts of partial and complete observables, suggested in \cite{RovPartObs}, for canonical systems with an arbitrary number of constraints. As has been shown, some of the ideas involved here have already appeared earlier, for instance in Kuchar's Bubble Time Formalism \cite{bubble}, Torre's work \cite{Torre1} or as the concept of a gauge invariant extension of a gauge restricted functions \cite{henneaux, smolin}.

Throughout this work we made global and local assumption on the clock variables and their properties under a general gauge transformation. Of course such assumptions are very difficult to verify in more complicated systems. It may also happen that there do not exist clock variables satisfying these assumptions. For instance, the choice of clock variables corresponds to a choice of a gauge and there exist in some systems obstructions to find a gauge satisfying the conditions mentioned at the beginning of section \ref{gauge}. Also, Kuchar's work \cite{bubble} was motivated by the hope to find a canonical transformation for the gravitational phase space variables, such that gravity can be seen as an already parametrized theory. But Torre showed that this is at least globally not possible, see \cite{Torre}.

So one of the future tasks will be, to weaken the global assumptions. A starting point for this could be the investigation of the convergence properties of the formal series (\ref{4pde29}). If this series converges (which has to be checked case by case), it would give a definition of Dirac observables, which involves only local information. In connection with these questions it would be interesting to know how the methods developed here apply to chaotic systems.

Moreover we showed in section \ref{partinv} that it is possible to work with a ``degenerate'' set of clock variables, which does not provide a perfect parametrization of the gauge orbit, as long as the partial observable $f$ has the same kind of degeneracy along the gauge orbit. So it might be not necessary to find clock variables, which lead to good gauges in the sense of \cite{henneaux}.

A crucial point in our methods is, that it is possible to obtain a weak abelianization of the constraints, moreover the clock variables evolve linearly with respect to these abelianized constraints. Effectiviely we have choosen a very convenient basis of the constraint set. This might be also helpful for the quantization of constraint systems, see \cite{master} for a related idea. 

Since there is not much hope to obtain exact expressions for Dirac observables (for gravity), we think that it is important to develop a perturbative treatment. For instance in \cite{torreanderson} (and references therein) it is explained, that Dirac observables for general relativity neseccarily contain spatial derivatives of infinitly high order. For the complete observables, these will be generated through the iterated Poisson brackets, that appear in the formal power series for a complete observable. 

One starting point for a perturbative treatment for Dirac observables could be the formal solution (\ref{4pde29}). Moreover one can hope to apply the perturbation theory for one constraint (i.e. the perturbation theory for ordinary ``classical mechanics'' systems, where the Hamiltonian is viewed as a constraint), if one uses the weakly abelianized constraints. Here the first difficulty one has to solve is to invert the matrix $A_{jk}$. To facilitate this task, it is important to make a clever choice of the clock variables. 

The concepts of partial and complete observables show, that Dirac observables involve the dynamical information of the system (if the constraints describe also the dynamics, as is the case for general relativity). Hence to obtain Dirac observables, it is necessary to solve certain (or in some cases all) aspects of the dynamics of the theory. Here we hope that the direct physical interpretation of the complete observables in terms of the partial observables facilitates the choice of the partial observables, one wants to begin with in the construction of Dirac observables. 

The main reason for the search of Dirac observables in gravity is, that they are needed as a crucial input in a canonical quantization. Here it is important to know the Poisson algebra of the Dirac observables, since during quantization one seeks for a representation of this Poisson algebra on a physical Hilbert space. First steps to obtain this Poisson algebra were undertaken in section \ref{gauge}. Of course, it is important to clarify at least the existence of Dirac observables.

Moreover, since complete observables are gauge invariant extensions of gauge restricted functions, one can compare the quantization based on gauging with a constraint quantization of the ungauged theory. This could clarify, to which extend we can rely on quantization schemes based on gauging.

\vspace{1cm}

{\Large {\bf Acknowledgements}}

I thank Thomas Thiemann for many helpful comments and discussions and for proposing the issue of Dirac observables as the topic of my PhD-thesis. This work was supported by a grant from the German National Merit Foundation and by a grant from NSERC of Canada to the Perimeter Institute for Theoretical Physics.

\end{document}